\documentclass[showpacs,preprintnumbers,amsmath,amssymb]{revtex4}
\usepackage{graphicx}
\usepackage{dcolumn}
\usepackage{bm}

\begin{document}

\allowdisplaybreaks

\preprint{
Preprint TTK-10-23%,\, 
%yymm.nnnn [hep-ph]
}

\title{Boltzmann  hierarchy for the cosmic microwave background at second 
order including photon polarization}
\author{M. Beneke and C.~Fidler}%
%\email{fidler@physik.rwth-aachen.de}
\affiliation{
Institut f\"ur Theoretische Teilchenphysik und Kosmologie, 
RWTH Aachen University, 
D - 52056 Aachen, Germany}

\date{\today}

\begin{abstract}
\noindent
Non-gaussianity and $B$-mode polarization are particularly interesting
features of the cosmic microwave 
background, as -- at least in the standard model of 
cosmology -- their only sources to first order in cosmological
perturbation theory are primordial, possibly generated during
inflation. If the primordial 
sources are small, the question arises how large is the
non-gaussianity and $B$-mode background induced in second-order 
from the initially gaussian and scalar perturbations. 
In this paper we derive the Boltzmann 
hierarchy for the microwave background photon phase-space distributions 
at second order in cosmological perturbation theory including the 
complete polarization information, providing the basis for 
further numerical studies. As an aside we note that 
the second-order collision term contains new sources 
of $B$-mode polarization and that no polarization persists in the
tight-coupling limit.
\end{abstract}

\maketitle
\arraycolsep0.07cm

%%%%%%%%%%%%%%%%%%%%%%%%%%%%%%%%%%%%%%%%%%%%%%%%%%%%%%%%%%%%%%%%%%%%%%%%%
\section{Introduction}

\noindent
The anisotropies of the cosmic microwave background provide an
abundant source of information on the early history of the universe 
unrivaled in precision. Since most of the photons originate from 
the time of decoupling, when the inhomogeneities of the universe 
were small, the anisotropies should be well described in linear 
perturbation theory around the Friedmann-Robertson-Walker background. 
Temperature anisotropies~\cite{Smoot:1992td,Larson:2010gs} and 
$E$-mode polarization 
anisotropies~\cite{Kovac:2002fg,Bennett:2003bz,Larson:2010gs} 
have been detected, and are found in agreement with the Standard
Model, in which the source of the anisotropies consists of a gaussian,
adiabatic and nearly scale-invariant spectrum of primordial 
density perturbations.

The polarization pattern of the background radiation is of great 
interest for the following reason. In contrast to the 
$E$ mode, $B$-mode polarization is not sourced by scalar density 
perturbations in the linear order. Thus, a detection of $B$-mode 
polarization would point directly to primordial vector or, more
likely, tensor fluctuations (gravitational waves) in a very early 
phase of the cosmological evolution. So far, however, $B$-mode 
polarization has not been observed, which together with the shape 
of the temperature perturbation spectrum indicates some suppression of 
tensor relative to scalar perturbations. Similarly, deviations 
from gaussian statistics constrain inflation models and are 
therefore intensively investigated (see, e.g., 
the reviews \cite{Bartolo:2004if,Liguori:2010hx}).

The absence of non-gaussianity and $B$-mode polarization when the 
primordial fluctuations are purely gaussian and scalar holds, 
however, only in linear perturbation theory. Thus, if a small 
non-gaussian or $B$-mode signal is observed, the question 
arises whether its origin is truly primordial, or whether it might 
be a second-order effect. While such an effect would naturally be 
expected at tensor-to-scalar ratios of order $10^{-5}$ (or 
$f_{\rm NL}\sim 1$ for non-gaussianity), which is 
the size of perturbations in the microwave background, only a full 
second-order calculation can tell whether there are no 
enhancements. Such enhancements can reach a level relevant 
to observations, since the planned CMBPol experiment is 
sensitive to tensor-to-scalar ratios of order 
$10^{-3}$~\cite{Baumann:2008aq}. Indeed, several second-order sources 
of $B$-mode polarization are already known. The most important is the 
weak-lensing effect, reviewed in \cite{Lewis:2006fu}, which converts 
$E$ polarization to $B$ polarization as the photons travel through the
inhomogeneous universe~\cite{Zaldarriaga:1998ar}. 
The inhomogeneities and the $E$-mode
polarization are both at least of first order, so the
resulting effect is at least of second order in perturbation theory. 
Weak lensing becomes large at small scales, and at large values of 
the perturbation wave-vector $\bm{k}$ perturbation series breaks
down. The usual treatment of weak lensing therefore avoids cosmic 
perturbation theory by considering the small deflection angles of the 
photon trajectories. Another effect that has been estimated is 
$B$-mode polarization from gravitational time delay \cite{Hu:2001yq} 
and from sources proportional to second-order vector and tensor metric
perturbations, which are themselves generated from the 
product of scalar perturbations~\cite{Mollerach:2003nq}. However, 
a full treatment of $B$-mode polarization at second-order is still 
missing, even at the qualitative level, since previous 
second-order calculations considered the collision term 
\cite{Dodelson:1993xz} and radiation transfer function
\cite{Bartolo:2006cu} for unpolarized radiation only. In this 
paper, we derive the complete Boltzmann hierarchy at second order 
under the assumption that there are no first-order vector and 
tensor perturbations 
by extending the results of \cite{Dodelson:1993xz,Bartolo:2006cu} to the 
photon polarization density matrix.
This allows us to identify all 
sources of $B$-mode polarization at this order. The polarized
equations are presented in a form suitable for numerical evaluation. 
Numerical results for $B$ polarization will be presented in a 
follow-up paper \cite{BFK10}. Numerical results on non-gaussianity
based on second-order Boltzmann equations have appeared recently in 
\cite{Bartolo:2010qu,Pitrou:2010sn}.

Most of the results of this paper have been obtained in the thesis 
work~\cite{FidlerThesis}. In the meantime, the polarized second-order 
Boltzmann equations have been derived independently 
in \cite{Pitrou:2008hy}. Our result is 
derived in a different formalism, allowing for an independent 
check of the results. We provide expressions 
for the Boltzmann hierarchy pertaining to 
the phase-space distribution functions 
not integrated over frequency, which have not been given explicitly before. 
In addition, we include a self-contained derivation 
of the polarized collision term from the quantum-mechanical time 
evolution of the photon density matrix, which differs from the 
collision term used in~\cite{Pitrou:2008hy}. 

The outline of the paper is as follows. In the remainder of this
section we set up our index conventions. Next, in 
Section~\ref{sec:boltzmann} we derive the Boltzmann equation from 
the quantum-mechanical time evolution of the photon density matrix. 
Its expansion to second order is presented in Sections~\ref{sec:expansion} 
and \ref{sec:expcollision}, first for the propagation of polarized 
radiation (the ``left-hand side'' of the Boltzmann equation), 
then for the collision term (the ``right-hand side''). These sections 
are rather technical, and further technical details are collected in 
Appendix~\ref{app:appendix}. The main result, the Boltzmann 
hierarchy for the second-order intensity and photon polarization 
phase-space distributions, is summarized in the separate 
Section~\ref{sec:eqsummary}. A full analysis of these equations is 
beyond the scope of this paper. However, in
Section~\ref{sec:discussion} we discuss the sources for $B$-mode 
polarization, including a new source in the collision term that 
converts intensity directly into $B$-mode polarization. We also 
analyze the tight-coupling regime. While we reproduce the 
presence of a second-order intensity 
quadrupole already found from the unpolarized 
equations~\cite{Bartolo:2006fj}, we do not confirm the effect
discussed in~\cite{Bartolo:2007dh}, which is based on $B$-mode 
generation from an $E$-mode and intensity quadrupole in
tight-coupling. Section~\ref{sec:conclusion} summarizes our 
conclusions. 

\subsection{Index and metric conventions}
\label{subsec:conventions}

\noindent
General coordinate indices will be denoted by Greek letters 
$\mu,\nu,\ldots$ ranging from 0 to 3, 
indices referring to tensors in the local 
inertial frame (tetrad frame) by capital Latin letters
$A,B, \ldots = 0,1,2,3$ from the beginning of the alphabet. Spatial 
indices, ranging from 1 to 3, in the tetrad frame are assigned 
Latin letters $i,j,\ldots$. We also need small Latin letters 
$a,b,\ldots =1,2$ to denote the basis of polarization vectors. 
The signature convention for the space-time metric is $(+,-,-,-)$. 
Spatial indices in the tetrad frame are contracted with the 
three-dimensional Euclidean metric and no distinction is made 
between upper and lower spatial indices. With this convention 
$v^A w_A = v^0 w^0 - v^i w^i = v^0 w^0 - v^i w_i$ etc.
In this paper we assume that the background universe is flat. 
We then also use Latin letters $i,j,\ldots$ to denote the 
spatial general coordinate indices with the same convention 
regarding their contraction. In general, it will be clear from the 
context whether $i,j,\ldots$ refers to the tetrad or general 
coordinate system. Since confusion might arise for the momentum, 
we denote the covariant momentum $d x^\mu/d\lambda$ by 
capital $P^\mu$, related to the momentum $p^A$ in the local 
inertial frame by 
$P^\mu = [e_A]^\mu p^A$ with $p^A$.

The perturbed flat-space Robertson-Walker metric with conformal time denoted 
by $\eta$ and coordinates 
$x^\mu = (\eta, x^i\,)$ is parameterized as  
\begin{eqnarray} 
ds^2 &=& a^2\,\Big((1+2A)\,d\eta^2 +2B_i \,d\eta dx^i 
-[(1+2D)\,\delta_{ij}+2E_{ij}]\,dx^idx^j\,\Big),
\label{metricpar}
\end{eqnarray}
where $a(\eta)$ denotes the scale factor.
The space-time dependent perturbations 
$X=A,D,B_i,E_{ij}$ will be expanded into 
first-order, second-order etc. terms according to 
$X=X^{(1)}+X^{(2)}+\ldots$. We assume that the vector and 
tensor perturbations contained in $B_i$, $E_{ij}$ 
are smaller than the scalar perturbations, 
so we formally treat them as second order.

We choose the conformal Newtonian 
gauge such that $B_i$ is a transverse vector and 
$E_{ij}$ a transverse, traceless tensor. In this case, when 
there are only scalar perturbations at first order, we have 
$B_i^{(1)}=E_{ij}^{(1)} =0$.
Then the tetrad components $[e_A]^\mu$ are given to second-order by
\begin{eqnarray}  
[e_0]^0 &=& \frac{1}{a}\,\Big(1-A^{(1)}-A^{(2)} + 
\frac{3}{2} {A^{(1)}}^2 
+\frac{1}{2}U_i^{(1)}U_i^{(1)}\,\Big),  
\nonumber\\[0.2cm]
[e_0]^i &=&  \frac{U_i}{a},
\nonumber\\[0.2cm]
[e_k]^0 &=&  \frac{1}{a}\left(U_k^{(1)}+U_k^{(2)}-B_k^{(2)}
+(D^{(1)}-A^{(1)}) \,U^{(1)}_k\right),
\nonumber\\[0.2cm]
[e_k]^i &=& \frac{1}{a}\,\Big(\delta_{ik} 
\,\Big(1-D^{(1)}-D^{(2)}+\frac{3}{2}{D^{(1)}}^2\,\Big) 
- E_{ik}^{(2)} - \frac{1}{2} U_i^{(1)} U_k^{(1)} \Big). 
\label{deftetrad}
\end{eqnarray}
We denote by $[e^A]_\mu$ the inverse of the tetrad, such that 
$[e^A]_\mu [e_B]^\mu = \delta^A_B$ and $[e^A]_\mu [e_A]^\nu 
=\delta^\nu_\mu$.
The local inertial frame can be parameterized in terms of the observer 
three-velocity $U_i$ and a set of angles $\theta_k$, which define 
the orientation of the local inertial coordinate axes relative 
to those of $x^i$. Above we have aligned the local coordinate axes 
with those of the general coordinate system to set the 
angles $\theta_k=0$. In the following we also choose the observer 
rest frame $U_i=0$, which coincides to first order with another 
common frame choice $U_i=B_i$.

\section{Boltzmann equation for the polarization density matrix}
\label{sec:boltzmann}

\noindent
In this section we briefly review notation and definitions applying 
to photon polarization. We then derive an expression for the
propagation and collision term in the Boltzmann equation for the 
polarization density matrix, which serves as the starting point for 
the expansion to second order in perturbations.

\subsection{Photon polarization phase-space distribution}
\label{subsec:photonpol}

\noindent
We assume that the polarized radiation ensemble can be described by a 
single-particle phase-space distribution 
matrix $\hat f_{\mu\nu}(x^\lambda,q^i)$, 
such that $\hat\epsilon^\mu \hat\epsilon^{\nu\star} 
\hat f_{\mu\nu}(x^\lambda,q^i)\,d^3p/(2\pi)^3$ denotes the 
number density of photons with momentum $\bm{p}$ and 
polarization $\hat \epsilon^\mu$. 
We regard $\hat f_{\mu\nu}$ as a function of the comoving 
momentum $q^i = a p^i$. The unperturbed Bose-Einstein 
distribution $\hat f_{\mu\nu}^{(0)}$ is then 
independent of conformal time in the expanding homogeneous 
universe. 

The phase-space distribution is a Hermitian matrix, 
related to the expectation value 
$\langle A_\mu(x) A_\nu(y)\rangle$ of the radiation field. We adopt 
Lorenz gauge $A^\mu_{\phantom{\mu};\mu} = 0$ for the photon field. 
It then follows that 
\begin{equation}
p^\mu \hat f_{\mu\nu}(x^\lambda,q^i) = 
p^\nu \hat f_{\mu\nu}(x^\lambda,q^i)=0, 
\end{equation}
and that 
$\hat f_{\mu\nu}$ is parallel-transported in the 
absence of collisions. Thus
\begin{equation}
\frac{\cal D}{{\cal D}\lambda} \hat f_{\mu\nu} = 
\hat C_{\mu\nu}[\hat f\,].
\label{boltzmann_fhat}
\end{equation}
Here ${\cal D}/{\cal D}\lambda$ denotes the covariant 
derivative along a photon trajectory 
$x^\mu(\lambda)$, and $\hat C_{\mu\nu}[\hat f\,]$ is the collision term. 
The phase-space distribution distribution  in the 
local inertial frame is related to $\hat f_{\mu\nu}$ by 
\begin{equation}
\hat f^{\mu\nu} = [e_A]^\mu [e_B]^\nu \hat f^{AB},
\end{equation}
where $[e_A]^\mu$ are the space-time dependent 
tetrad vectors.

The phase-space distribution  matrix $\hat f_{\mu\nu}$ is not unique, 
since Lorenz gauge allows the gauge transformations   
$\hat f_{\mu\nu} \to \hat f_{\mu\nu} + \alpha_\mu P_\nu +\beta_\nu
P_\mu$ with arbitrary $\alpha_\mu$, $\beta_\nu$. To obtain a 
physical distribution function, we decompose  
the photon four-momentum into 
\begin{equation}
P^\mu = E \,[e_0]^\mu - [e_i]^\mu p^i = E \,(u^\mu - n^\mu),
\end{equation}
where $u^\mu = [e_0]^\mu$ is the four-velocity of the locally inertial 
observer, $E$ the energy of the photon as seen by this observer, 
and 
\begin{equation}
n^\mu = [e_i]^\mu\,\frac{p^i}{E}
\end{equation}
the photon three-momentum direction, which satisfies $u_\mu n^\mu=0$ 
and $n_\mu n^\mu=-1$. We define 
\begin{equation}
p_{\mu\nu} = -g_{\mu\nu} + u_\mu u_\nu - n_\mu n_\nu,
\end{equation}
which projects on the components transverse to the observer velocity 
and photon direction:
\begin{equation}
u^\mu p_{\mu\nu} = u^\nu p_{\mu\nu} = 
n^\mu p_{\mu\nu} = n^\nu p_{\mu\nu} = 0.
\end{equation} 
We now define the physical phase-space distribution matrix
\begin{equation}
f_{\mu\nu} = p_{\mu}^{\phantom{\mu}\mu^\prime}
p_{\nu}^{\phantom{\nu}\nu^\prime} 
\hat f_{\mu^\prime\nu^\prime},
\label{deff}
\end{equation}
which is orthogonal to $P^\mu$, $u^\mu$ and $n^\mu$, and contains 
no residual gauge ambiguity. The corresponding projected 
distribution function in the observer rest-frame (local inertial 
frame) is effectively a three-by-three matrix, since 
$f^{00} = f^{0i}=f^{i0}=0$ in this frame. It is orthogonal to the photon 
propagation direction $n^i$ as expected for the tensor describing 
the two physical transverse polarizations.

Applying  ${\cal D}/{\cal D}\lambda$ to (\ref{deff}), we obtain 
from (\ref{boltzmann_fhat})
\begin{equation}
p_\mu^{\phantom{\mu}\mu^\prime}p_\nu^{\phantom{\nu}\nu^\prime} 
\frac{\cal D}{{\cal D}\lambda} f_{\mu^\prime\nu^\prime} = 
p_\mu^{\phantom{\mu}\mu^\prime}p_\nu^{\phantom{\nu}\nu^\prime} 
\hat C_{\mu^\prime\nu^\prime}[\hat f\,] ,
\label{boltzmann_f}
\end{equation}
where we use that $p_{\mu\nu}$ and $\hat f_{\mu\nu}$ are 
orthogonal to $P^\mu$. This provides a closed set of equations for 
the physical phase-space distribution matrix provided the 
projected collision term $C_{\mu\nu}[\hat f\,] 
\equiv p_\mu^{\phantom{\mu}\mu^\prime}p_\nu^{\phantom{\nu}\nu^\prime} 
\hat C_{\mu^\prime\nu^\prime}[\hat f\,]$ on the right-hand side depends only 
on $f$ rather than $\hat f$. It will be convenient to choose 
a polarization basis consisting of two vectors $\epsilon^\mu_a$ 
($a=1,2$) orthogonal to $P^\mu$ and $u^\mu$, in terms of 
which 
\begin{equation}
f^{\mu\nu} = \sum _{a,b} f_{ab} \,\epsilon_a^{*\mu} \epsilon_b^\nu.
\label{deffAB}
\end{equation}

\subsection{Propagation of polarized photons}
\label{subsec:propagation}

\noindent
In the following we shall consider the Boltzmann equation 
for $f_{ab}$ in a polarization basis defined in the observer
rest-frame. To this end, we multiply (\ref{boltzmann_f}) by 
$\epsilon^\mu_a \epsilon^{*\nu}_b$, insert  (\ref{deffAB}), 
and calculate
\begin{equation}
\epsilon^\mu_a \epsilon^{*\nu}_b\,
p_{\mu\mu^\prime}p_{\nu\nu^\prime} 
\frac{\cal D}{{\cal D}\lambda} \,
 \sum _{c,d} \,f_{cd} \,
\epsilon_c^{*\mu^\prime} \epsilon_d^{\nu^\prime}.
\label{lhs1}
\end{equation}
Before giving the result, we need to discuss the choice of 
polarization basis. If $\hat \epsilon^\mu$ denotes the 
direction of the photon field-amplitude in Lorenz gauge, 
then $\hat \epsilon^\mu$ is parallel-transported along the 
photon path \cite{MTWbook}. This also holds for the projected 
vector $\epsilon^\mu = p^{\mu}_{\phantom{\mu}\mu^\prime}\, 
\hat \epsilon^{\mu^\prime}$ in the sense that for the polarization 
vector corresponding to a fixed photon momentum
$p^{\mu}_{\phantom{\mu}\nu} 
\,{\cal D} \epsilon^\nu/{\cal D}\lambda =0$. 
We could therefore use a basis of two 
parallel-transported polarization vectors 
$\epsilon_a^\mu$. In this case, the corresponding 
vectors $\epsilon^A_a$ in the observer rest-frame, given by 
$\epsilon_a^\mu = [e_A]^\mu \epsilon^A_a$, depend 
on the space-time point $x$ in addition to momentum $q$. The 
transverse polarization 
basis axes are continuously rotated as the photon propagates 
through space-time.

It will be convenient to instead choose rigid basis vectors 
$\epsilon^A_a$ in the observer rest-frame that do not depend 
on space-time but only on momentum. 
In this case the covariant derivative along 
the path acting on a polarization vector is 
\begin{equation}
\epsilon_{a\mu}^* \frac{{\cal D} \epsilon_c^\mu}{{\cal D}\lambda}
= [e^B]_\mu [e_A]^\mu_{\phantom{\mu};\nu} \,P^\nu \epsilon_{aB}\,\epsilon_c^A 
+ \epsilon_{aA} \frac{dq^i}{d\lambda} 
\frac{\partial \epsilon_c^A}{\partial q^i},
\end{equation}
where the first term on the right-hand side arises, since 
$\epsilon_c^\mu$ is not parallel-transported in the rigid basis, 
and the second is due to the fact that the photon momentum also 
changes along the path. The conditions 
$u_\mu \epsilon^\mu_a=P_\mu \epsilon^\mu_a=0$ require 
$\epsilon^0_a=0$ and $\bm{n}\cdot \bm{\epsilon}_a = 0$ 
in the observer rest-frame. We adopt a spherical coordinate 
system with 
\begin{eqnarray}
&& \bm{n} = (\sin\theta \cos\varphi, \sin\theta\sin\varphi,
\cos\theta)
\nonumber \\ 
&& \bm{e}_\theta = (\cos\theta \cos\varphi, \cos\theta\sin\varphi,
-\sin\theta)
\nonumber\\
&& \bm{e}_\varphi = (-\sin\varphi,\cos\varphi,0)
\end{eqnarray} 
the standard basis vectors on the sphere. The polarization basis 
is then taken to consist of the two circular polarization vectors
\begin{equation}
\bm{\epsilon}_\pm =
-\frac{1}{\sqrt{2}}(\bm{e}_\theta \pm  i\bm{e}_\varphi).
\label{def:polvectors}
\end{equation}
We also identify $x^0=\eta$ 
with conformal time. 

Evaluating (\ref{lhs1}) results in the Boltzmann equation
\begin{eqnarray}
\nonumber
&& \frac{\partial f_{ab}}{\partial\eta} +
\frac{1}{P^0} \frac{dx^i}{d\lambda}\frac{\partial f_{ab}}{\partial x_i}
+\frac{1}{P^0} \frac{d q^i}{d\lambda} 
\left(\frac{\partial f _{ab}}{\partial q^i} 
+\epsilon_{ak}\frac{\partial \epsilon_c^{k*}}{\partial {q}^i}
f_{cb} +\epsilon_{bk}^*
 \frac{\partial \epsilon_c^k }{\partial {q}^i} f_{ac}
\right)
\nonumber \\[0.2cm]
&&\hspace*{0cm}
+\,\,[e^i]_\mu\, [e_k]^\mu_{\phantom{\mu};\nu}\,\frac{P^\nu}{P^0} 
\left(\epsilon_{ai} \epsilon_c^{*k} f_{cb} +
\epsilon_{bi}^{*}\epsilon_c^k f_{ac}\right) = \frac{1}{P^0}\,C_{ab}[f],
\label{boltzmann} 
\end{eqnarray}
with $C_{ab}[f] = \epsilon_a^\mu \epsilon_b^{*\nu} C_{\mu\nu}[f]$
denoting the (projected) 
collision term in the polarization basis. Its expression 
is given in (\ref{scatteringterm}) below. The terms to 
the left of the equality sign in the second line would vanish had we 
chosen a basis of parallel-transported polarization vectors. The 
extra terms in the rigid basis are equivalent to similar terms 
that appear in \cite{DR78}.

For unpolarized radiation $f_{ab} = \delta_{ab} \,f$, and the terms 
in (\ref{boltzmann}) that depend on the polarization vectors
explicitly vanish. This follows from 
\begin{equation}
\epsilon_{ak}\frac{\partial \epsilon_b^{k*}}{\partial {q}^i}
+\epsilon_{bk}^* \frac{\partial \epsilon_a^k }{\partial {q}^i} 
= \frac{\partial}{\partial {q}^i} \left(\epsilon_{ak}\epsilon_{b}^{*k}
\right)=0,
\end{equation}
since $\epsilon_{ak}\epsilon_{b}^{*k} = \delta_{ab}$, and from
\begin{eqnarray}
&& [e^i]_\mu\, [e_k]^\mu_{\phantom{\mu};\nu} 
\left(\epsilon_{ai} \epsilon_b^{*k}+
\epsilon_{bi}^{*}\epsilon_a^k\right) 
= \epsilon_{ai} \epsilon_b^{*k} \left([e^i]_\mu\, [e_k]^\mu\right)_{;\nu} 
=0\,,
\end{eqnarray}
since $[e^i]_\mu\, [e_k]^\mu = \delta^i_k$. Thus (\ref{boltzmann}) 
reduces to the standard equation for unpolarized radiation for 
diagonal phase-space density matrices, as should be the case.

For the same reason, the explicitly polarization-vector dependent 
terms are at least of second order in perturbations around the
equilibrium distribution in the expanding homogeneous universe. 
This is due to the fact that $dq^i/d\lambda$ and 
$[e^i]_\mu\, [e_k]^\mu_{\phantom{\mu};\nu}$ are both 
first order in perturbations.  Hence to first order we may set
$f_{ab}$ in the polarization-vector dependent terms equal to the unperturbed 
distribution $f_{ab}^{(0)}$. But the 
unperturbed distribution is diagonal, so the terms vanish (at first
order) as shown above.

\subsection{Simplification at second order}
\label{subsec:simplification}

\noindent
We now show that the terms to 
the left of the equality sign in the second line of (\ref{boltzmann}) 
vanish even at second-order in perturbations, provided 
that there are no first-order vector and tensor perturbations. 
Thus, under these assumptions, there is no difference between 
the rigid and the parallel-transported polarization basis at second 
order. 

It follows from Section~\ref{subsec:propagation} that the second-order
contribution is the product of 
\begin{equation}
\epsilon_{ai} \epsilon_c^{*k} f_{cb}^{(1)} +
\epsilon_{bi}^{*}\epsilon_c^k f_{ac}^{(1)}
\label{eq:eps1}
\end{equation}
and 
\begin{eqnarray}
&& \left[[e^i]_\mu\, [e_k]^\mu_{\phantom{\mu};\nu}\right]^{(1)}\,
\left[\frac{P^\nu}{P^0}\right]^{(0)}
= \,\frac{1}{2} \left(\partial_i B_k^{(1)} - \partial_k B_i^{(1)}
\right)
- n_l 
\left(\partial_i E_{kl}^{(1)} - \partial_k E_{il}^{(1)}\right) + \ldots\,,
\label{eq:secondsimp1}
\end{eqnarray}
where the superscript in brackets indicates the order at which the 
expression is evaluated, and the ellipses denote terms proportional 
to $q_i$ or $q_k$ which vanish when contracted with the polarization 
vectors in (\ref{eq:eps1}) since $\bm{q}\propto\bm{n}$. 
To obtain this expression we used the 
tetrads from (\ref{deftetradfull}), which do not assume a particular 
gauge choice. 

If there are no first-order vector and tensor perturbations, $B_i^{(1)}$  
and $E_{ij}^{(1)}$ are zero in conformal Newtonian gauge, and 
expression (\ref{eq:secondsimp1}) immediately vanishes, leading to the desired 
simplification. More generally, in an arbitrary gauge $B_i^{(1)}$
can be expressed as the gradient of a scalar function in the absence 
of vector modes, hence the curl of $B_i^{(1)}$ 
appearing in (\ref{eq:secondsimp1}) is zero. Likewise, 
$E_{ij}^{(1)} = (\partial_i\partial_j - \delta_{ij}\partial^2)\,
E^{(1)}$ for some function $E^{(1)}$ in the absence of first-order 
vector and tensor modes. Then, 
\begin{equation}
n_l \left(\partial_i E_{kl}^{(1)} - \partial_k E_{il}^{(1)}\right)
= (n_i\partial_k-n_k\partial_i)\,\partial^2 E^{(1)},
\end{equation} 
which vanishes when contracted with the polarization 
vectors in (\ref{eq:eps1}), since $n_i \epsilon_{ai}=0$.

\subsection{Collision term}
\label{sec:collisionterm}

\noindent
To obtain the collision term $C_{AB}[f]$ we consider the 
quantum time evolution of the one-particle density matrix following 
the formalism developed in \cite{Sigl:1992fn} for neutrino 
flavour-mixing in a medium. The formalism was applied to 
photon polarization and Thomson scattering in \cite{Kosowsky:1994cy}. 
A more general treatment elucidating some of the approximations 
involved in the truncation of the hierarchy of 
$n$-particle density matrices implicit in this formalism 
can be found in~\cite{Nagirner:2001xd}.

In the local inertial frame with coordinates $\xi$ 
the photon field operator is 
expanded in the form 
\begin{equation}
\bm{A}(\xi) =\sum _{a=\pm} \int \!\frac{d^3\bm{p}}{(2\pi)^3 2 p^0} 
\left(e^{-i p\cdot \xi} a_a(p) \bm{\epsilon}_a(p) +
e^{i p\cdot \xi} a_a^\dagger(p) \bm{\epsilon}^*_a(p)
\right).
\end{equation}
We choose the two circular polarization 
vectors $\bm{\epsilon}_\pm$ as basis vectors. The creation 
and annihilaton operators statisfy the standard commutation relation
\begin{eqnarray}
\big[a_a(p),a^\dagger_b(p^\prime)\big] &=& \delta_{ab} \,(2\pi)^3 2 p^0 
\delta^{(3)}(\bm{p}-\bm{p}^\prime)
\equiv \delta_{ab} \,\delta(p-p^\prime).
\end{eqnarray}

The one-particle density matrix is defined by the expectation 
value $\langle a_b^\dagger(p)a_a(p^\prime)\rangle$.
Spatial homogeneity implies that 
\begin{equation}
\langle a_b^\dagger(p)a_a(p^\prime)\rangle = 
\delta(p-p^\prime) \,\rho_{ab}(t,p).
\label{rhodef}
\end{equation}
We identify $\rho_{ab}(t,p)$ with the phase-space distribution 
function $f_{ba}(x^\lambda,q^i=ap^i)$. Indeed, since the number operator 
is 
\begin{equation}
\hat N =  \sum _{a=\pm} \int \!\frac{d^3\bm{p}}{(2\pi)^3 2 p^0} 
\,a_a^\dagger(p)a_a(p),
\end{equation}
we obtain from (\ref{rhodef})
\begin{equation}
N = \langle \hat N\rangle = V \int d^3\bm{p}\, \,\mbox{tr} \,\rho(t,p),
\end{equation}
confirming the interpretation of $\rho_{ab}(t,p)$ as phase-space 
polarization density matrix. The spatial dependence of 
$f_{ba}(x^\lambda,q^i)$ can be neglected for the calculation of 
the collision term, since each scattering event is local on the 
cosmological scales over which  $f_{ba}(x^\lambda,q^i)$ varies. The
flip in the order of polarization indices follows from the definitions
(\ref{deffAB}), (\ref{rhodef}) and the fact that
$\sum_a a_a(p)\bm{\epsilon}_a(p)$ is independent of the choice of
polarization basis. 

The time evolution of the density matrix is obtained 
from the Heisenberg equation for the operator 
$D_{ab}(p) = a_b^\dagger(p)a_a(p)$. Starting from 
\begin{equation}
\frac{d}{dt}\,D_{ab} = i\left[H,D_{ab}\right], 
\end{equation}
going to the interaction picture and splitting the Hamiltonian 
into the free and interaction part $H_I$, we obtain to second 
order in the interaction \cite{Sigl:1992fn}
\begin{eqnarray}
&&2 p^0 (2\pi)^3 \delta^{(3)}(0) \,\frac{d}{dt} \rho_{ab}(t,p) 
= i \langle \left[H_I(t),D_{ab}(t,p)\right]\rangle
- \int_0^tdt^\prime \,\langle \left[H_I(t-t^\prime),
\left[H_I(t),D_{ab}(t,p)\right]\right]\rangle.
\label{siglraffelt}
\end{eqnarray}
If $H_I$ were the electron-photon interaction of quantum
electrodynamics $H_{\rm QED}$, we would have to expand to the fourth order in 
the interaction to recover the Compton scattering collision term. 
Instead we derive an effective Compton scattering interaction 
vertex assuming that the electron propagates freely between the 
two elementary electron-photon interactions in the Compton process.  
Thus we define $H_I(t)$ through the relation
\begin{eqnarray}
(-i) \int dt \,H_I(t) &=&
\frac{(-i)^2}{2} \int d^4 x d^4 y \,
T(H_{\rm QED}(x) H_{\rm QED}(y))\qquad
\end{eqnarray}
with the understanding that a pair of electron fields is 
contracted in the expression on the right-hand side. After a short 
calculation we obtain  \cite{Kosowsky:1994cy}
\begin{eqnarray}
H_I(t) &=& \sum_{a,a^\prime,s,s^\prime} 
\int\,[dp][dp^\prime] [dq][dq^\prime]\,
(2\pi)^3\delta^{(3)}(\bm{q}^\prime+\bm{p}^\prime-\bm{q}-\bm{p})\,
e^{it (q^{0\prime}+p^{0\prime}-q^0-p^0)}
\nonumber\\
&&\,\times M(pa;qs\to p^\prime a^\prime;q^\prime s^\prime) 
\, \alpha^\dagger_{s^\prime}(q^\prime)a^\dagger_{a^\prime}(p^\prime) 
a_a(p) \alpha_s(q).
\end{eqnarray}
Here $\alpha$, $\alpha^\dagger$ denote electron annihilation and
creation operators, and $[dp] = d^3\bm{p}/((2\pi)^3 2 p^0)$ is 
the phase-space integration measure. The 
matrix element for the $\gamma(p,a)+e^-(q,s)\to \gamma(p^\prime,a^\prime)+
e^-(q^\prime, s^\prime)$  Compton scattering process reads 
\begin{equation}
M(pa;qs\to p^\prime a^\prime;q^\prime s^\prime) = 
e^2\,\bar u(q^\prime,s^\prime) 
\left[
\not\!\epsilon_{a^\prime}^{*}(p^\prime)
  \frac{\not\!q\,+\not\!p+m_e}{(q+p)^2-m_e^2} \not\!\epsilon_a(p) + 
\not\!\epsilon_{a}(p)
  \frac{\not\!q\,-\not\!p^\prime+m_e}{(q-p^\prime)^2-m_e^2} 
  \not\!\epsilon_{a^\prime}^*(p^\prime) 
\right] u(q,s)
\label{comptonME}
\end{equation} 
We note that 
\begin{equation}
M(pa;qs\to p^\prime a^\prime;q^\prime s^\prime) = 
M^*( p^\prime a^\prime;q^\prime s^\prime \to pa;qs).
\label{ccM}
\end{equation}
To avoid confusion let us also note that in this subsection $q$ 
stands for an electron momentum and not for the comoving photon momentum.

The first-order term  $\langle \left[H_I(t),D_{ab}(t,p)\right]\rangle$
in (\ref{siglraffelt}) involves the forward Compton scattering 
matrix element, and it is straightforward to show that this term 
vanishes. The second-order term is more complicated. It results in 
expectation values of four photon annihilation and creation 
operators, since the interaction generates correlations. To proceed 
we assume that $n$-particle correlations can be expressed in terms 
of one-particle correlations, such that, for example 
\begin{eqnarray}
\langle a^\dagger_{a^\prime}(q^\prime) a_a(q) 
a^\dagger_{b^\prime}(p^\prime) a_b(p)\rangle 
&=& \delta(q-p^\prime) \delta_{a b^\prime} 
\langle a^\dagger_{a^\prime}(q^\prime) a_b(p)\rangle +
\langle a^\dagger_{a^\prime}(q^\prime) a^\dagger_{b^\prime}(p^\prime) 
a_a(q) a_b(p)\rangle 
\nonumber\\[0.2cm]
&&\hspace*{-3cm} \,\longrightarrow\,
\delta(q-p^\prime) \delta_{a b^\prime} 
\langle a^\dagger_{a^\prime}(q^\prime) a_b(p)\rangle +
\langle a^\dagger_{a^\prime}(q^\prime) a_a(q) \rangle
\langle a^\dagger_{b^\prime}(p^\prime) a_b(p)\rangle +
\langle a^\dagger_{a^\prime}(q^\prime) a_b(p) \rangle
\langle a^\dagger_{b^\prime}(p^\prime) a_a(q)\rangle 
\nonumber\\[0.3cm]
&& \hspace*{-3cm} = \,
\delta(q-p^\prime) \delta(q^\prime-p)\rho_{ba^\prime}(p) 
\left[\delta_{a b^\prime} +\rho_{a b^\prime}(q)\right] 
 + 
\delta(q-q^\prime) \delta(p-p^\prime)\rho_{a a^\prime}(q) 
\rho_{bb^\prime}(p) .
\end{eqnarray}
This amounts to the assumption that on average multi-particle correlations 
built up in a collision decay rapidly in the time interval before 
the next collision.
The corresponding expressions for the electrons are simpler, since we 
further assume that the electrons are unpolarized and that their
phase-space density $g_e(q)$ is sufficiently small 
for quadratic terms in $g_e$ to be negligible. Thus 
\begin{eqnarray}
&& \langle \alpha^\dagger_{s^\prime}(q^\prime) \alpha_s(q) 
\alpha^\dagger_{r^\prime}(p^\prime) \alpha_r(p)\rangle 
\longrightarrow \,
\delta(q-p^\prime)\delta_{r^\prime s}
 \langle \alpha^\dagger_{s^\prime}(q^\prime) \alpha_r(p)\rangle 
= \delta(q-p^\prime)\delta(q^\prime-p)\delta_{s r^\prime} 
\delta_{r s^\prime} \,\frac{1}{2} g_e(q^\prime).
\quad
\end{eqnarray}
Note that  $g_e(q)$ is the density summed over both electron 
polarizations. 
After working out the expectation value of the second-order term in  
(\ref{siglraffelt}) one ends up with the time integral
\begin{equation}
\int_0^t dt^\prime \,e^{\pm i t^\prime (q^{\prime 0}+p^{\prime 0} 
-q^0-p^0)}.
\end{equation}
If the interaction time-scale is much shorter than the average time 
between collisions the upper limit may be taken to infinity and
supplying the appropriate $i\epsilon$ prescription, we obtain 
\begin{equation}
\pm i\,\mbox{PV}\,\frac{1}{q^{\prime 0}+p^{\prime 0} 
-q^0-p^0} + \pi \delta(q^{\prime 0}+p^{\prime 0} 
-q^0-p^0).
\end{equation}
The imaginary principal-value term should be discarded, since it 
corresponds to a self-energy contribution. Putting everything 
together, substituting $\rho_{ba} \to f_{ab}$ in the last step, 
we obtain from (\ref{siglraffelt}) 
\begin{equation}
2 p^0 \frac{d}{dt} f_{ab}(p)  = 2 C_{ab}[f], 
\end{equation}
where the collision term is given by
\begin{eqnarray}
C_{ab}[f] &=& \frac{1}{4} \,\int
\frac{d\bm{p}^\prime}{(2\pi)^3 2 p^{\prime 0}}
\frac{d\bm{q}}{(2\pi)^3 2 q^{0}}
\frac{d\bm{q}^\prime}{(2\pi)^3 2 q^{\prime 0}}\,
(2\pi)^4\delta^{(4)}(q+p-q^\prime-p^\prime)\,
|\overline{M}\,|^2_{\lambda\lambda^\prime;\omega\omega^\prime}
\nonumber\\[0.2cm]
&& \times\bigg\{g_e(q^\prime)
f_{\lambda^\prime\omega^\prime}(p^\prime)
\Big[\delta_{a\lambda}(\delta_{\omega b}+f_{\omega b}(p))+
\delta_{\omega b}(\delta_{a\lambda}+f_{a\lambda}(p))\Big]
\nonumber\\
&& \hspace*{0.5cm} -\,g_e(q)\Big[\delta_{a \lambda} f_{\omega b}(p) +
\delta_{\omega b} f_{a \lambda}(p)\Big]
(\delta_{\lambda^\prime\omega^\prime }+
f_{\lambda^\prime\omega^\prime}(p^\prime))\bigg\}.
\label{scatteringterm}
\end{eqnarray}
Here we introduced the electron-spin averaged square of the 
Compton amplitude
\begin{equation}
|\overline{M}\,|^2_{\lambda\lambda^\prime;\omega\omega^\prime}
=  \frac{1}{2} \sum_{s,s^\prime}
M(p\lambda;qs\to p^\prime \lambda^\prime;q^\prime s^\prime) 
M^*(p \omega;q s \to p^\prime\omega^\prime;q^\prime s^\prime).
\label{msquared}
\end{equation}
The collision term (\ref{scatteringterm}) for the polarized phase-space 
density is the expression that must be 
used on the right-hand side of the Boltzmann equation (\ref{boltzmann}). 
It takes an intuitive form with a gain and loss term and the expected Bose 
enhancement factors. Taking the trace in $ab$, and averaging the 
matrix element over polarizations, we recover the standard
unpolarized collision term. Eq.~(\ref{scatteringterm}) differs 
from \cite{Kosowsky:1994cy}, where it is stated that the terms 
quadratic in the photon phase-space density cancel exactly in the 
evaluation of the double commutator in (\ref{siglraffelt}). It 
also differs from the collision term used in \cite{Pitrou:2008hy}, 
which is based on \cite{Portsmouth:2004ee}. The differences are 
located in the structure of the loss term from
\cite{Portsmouth:2004ee} and the Bose enhancement factors 
added in \cite{Pitrou:2008hy}. The loss term 
in~\cite{Portsmouth:2004ee} is not derived as in the present paper 
but based on a certain ansatz, which is checked for 
initial and final pure photon polarization states, 
and then argued to hold in general 
due to the superposition principle. However, the loss term ansatz 
in~\cite{Portsmouth:2004ee} is non-linear in the phase-space 
distribution invalidating the superposition principle, 
and we suspect that this leads to the discrepancy 
with our result. Nevertheless, it turns out that the differences 
do not affect the final result in Section~\ref{sec:eqsummary} 
below after the expansion to second
order, at least for the 
frequency-integrated phase-space distributions considered 
in  \cite{Pitrou:2008hy}. The reason for this is the simple
polarization dependence of the Thomson scattering cross section 
and the fact that the terms quadratic in the photon phase-space
densities will be seen to not contribute to the second-order 
equations for the frequency-integrated distributions. Differences 
between the present calculation and  \cite{Pitrou:2008hy} from 
the form of the collision term would however be expected 
at the next order.

%[Add discussion. Discuss why flat-space derivation is valid with ] 

\subsection{Fourier transformation and multipole expansion}
\label{subsec:ftandlm}

\noindent
It is more convenient for the perturbation expansion to work with 
Fourier-transformed and multipole-expanded functions. We define 
\begin{equation}
A(\bm{x}) = \int \frac{d^3\bm{k}}{(2\pi)^3} \,
e^{i \bm{k}\cdot\bm{x}}\,A(\bm{k}).
\end{equation}
At second order we encounter products of functions, whose Fourier
transform is a convolution. Below we use the short-hand notation 
%\begin{widetext}
\begin{equation}
A(\bm{k}_1) B(\bm{k}_2) \equiv 
\int \frac{d^3\bm{k}_1}{(2\pi)^3} 
\int \frac{d^3\bm{k}_2}{(2\pi)^3} \,
(2\pi)^3\delta^{(3)}(\bm{k}-\bm{k}_1-\bm{k}_2) \,
A(\bm{k}_1) B(\bm{k}_2). 
\label{convolution}
\end{equation}

For the multipole representation we write the comoving momentum 
as $\bm{q} = q\bm{n}$ and then define
\begin{eqnarray}
&& f_{ab}(\eta,\bm{k},\bm{q}) =
\sum_{l=0}^\infty\sum_{m=-l}^l 
(-i)^l \sqrt{\frac{4\pi}{2 l+1}} \,f_{ab,lm}(\eta,\bm{k},q)\,
Y_{lm}^s(\bm{n}),
\label{lm1}\\
&& f_{ab,lm}(\eta,\bm{k},q) = 
i^l \sqrt{\frac{2 l+1}{4\pi}} \int d\Omega \,Y_{lm}^{s*}(\bm{n}) 
\,f_{ab}(\eta,\bm{k},q\bm{n}).
\label{lm2}
\end{eqnarray}
Here $Y_{lm}^s(\bm{n})$ denotes the spin-weighted spherical harmonic.
We collect the definitions and some basic relations for these
functions in appendix~\ref{app:ys}.
%\end{widetext}

We adopt the circular polarization basis (\ref{def:polvectors}) 
such that under a rotation of the coordinate system around the 
direction of photon propagation with rotation angle $\Delta \Psi$ the 
polarization basis vectors transform according to
\begin{eqnarray}
\bm{\epsilon}^\prime_{a = \pm} = e^{\pm i \Delta \Psi}
\bm{\epsilon}_{a = \pm},
\end{eqnarray}
i.e.~the circular polarization vectors $\bm{\epsilon}_\pm$ have spin 
$s=\pm 1$ as they should.
Since the polarization-basis independent phase-space distribution 
\begin{eqnarray}
\label{eq:PolTensor}
f^{ij} = \sum \limits_{ab} \epsilon^{i*}_{a}\epsilon^j_{b}f_{ab}
\end{eqnarray}
is invariant under basis rotations, 
it follows that $f_{++}$ and $f_{--}$ are spin-zero 
($s=0$) objects that do not transform, while 
\begin{eqnarray}
f'_{\pm \mp} = e^{\pm 2 i \Delta \Psi}f_{\pm \mp}.
\end{eqnarray}
Thus, $f_{+-}$ has spin 2 and $f_{-+}$ has spin $-2$. The corresponding 
values of $s$ must be used in (\ref{lm1}), (\ref{lm2}).

Instead of the phase-space densities of the photon helicity states,
one may also parameterize $f_{ab}$ in terms of the four real Stokes
parameters. The relation in the circular basis is 
\begin{equation}
f_{ab} = 
\left(\begin{array}{cc} f_{++} & f_{+-} \\ 
f_{-+} & f_{--}\end{array}\right) = 
\left(\begin{array}{cc} f_I-f_V & f_Q-if_U \\ 
f_Q+if_U & f_I+f_V\end{array}\right). 
\label{eq:Stokes}
\end{equation}
The multipole decomposition for the Stokes parameter distribution
functions $f_X$ reads
\begin{eqnarray}\nonumber
f_{I,lm} &=& i^l \sqrt{\frac{2l+1}{4\pi}}\int d\Omega \,
Y_{lm}^*(\bm{n}) f_I(\bm{n}), \\ \nonumber
f_{V,lm} &=& i^l \sqrt{\frac{2l+1}{4\pi}}\int d\Omega \,
Y_{lm}^*(\bm{n}) f_V(\bm{n}), \\ 
f_{E,lm} \pm if_{B,lm} &=& i^l \sqrt{\frac{2l+1}{4\pi}}\int d\Omega \,
Y_{lm}^{ \mp 2 *}(\bm{n}) \,[f_Q(\bm{n}) \pm i f_U(\bm{n})].
\label{eq:Multipol}
\end{eqnarray}
The quantity $f_I$ provides the photon density averaged over the two
helicity states, and $f_V$ is related to the degree of circular
polarization of the radiation plasma. We shall include $f_V$ in the 
set of second-order equations, but since there are no sources of 
circular polarization in the standard cosmological scenario, it is 
usually of little interest. Our main concern are the off-diagonal 
components of the photon phase-space density, which are decomposed in 
(\ref{eq:Multipol}) into the $E$ and $B$ polarization modes. 
The conversion between the two sets of phase-space distributions follows 
from 
\begin{equation}
f_{X,lm} = U_{X;[ab]} f_{ab,lm}.
\label{IVEBtransform}
\end{equation}
Interpreting $[ab]$ as a single index taking the values $++, --, +-,-+$
in this order, and with $X=I,V,E,B$, the matrix $U_{X;[ab]}$ and its 
inverse read
\begin{equation}
U_{X;[ab]} = \left(\begin{array}{rrrr}
\frac{1}{2} & \phantom{+}\frac{1}{2} & 0 & 0\\[0.2cm]
- \frac{1}{2} & \frac{1}{2} & 0 & 0\\[0.2cm]
0 &0 & \frac{1}{2} & \frac{1}{2}\\[0.2cm]
0 &0 & - \frac{1}{2i} & \frac{1}{2i}
\end{array}\right), 
\qquad\quad
U^{-1}_{[ab];X} = \left(\begin{array}{rrrr}
1 & -1 & 0 & 0\\[0.2cm]
1 & 1  & 0 & 0\\[0.2cm]
0 &0 & \phantom{+}1 & -i \\[0.2cm]
0 &0 & 1  & i
\end{array}\right). 
\label{Umatrices}
\end{equation}
We note the relations 
\begin{equation}
U^{-1}_{[ab];X} = U^{-1\,*}_{[ba];X} = 2 \,U^*_{X;[ab]}. 
\label{Urel}
\end{equation}

In terms of multipoles the momentum derivative terms 
in the first line of the Boltzmann equation (\ref{boltzmann}) 
can be written in a simple form. First, from (\ref{def:polvectors}) 
we calculate 
\begin{equation}
\epsilon_{bk}^*
\frac{\partial \epsilon_c^k }{\partial {q}^i} = 
\mp\frac{i}{q  \tan\theta} \,e_{\varphi i} \delta_{bc}
\end{equation}
where the upper (lower) sign holds for $b=c=+$ ($b=c=-$). 
Then, making use of (\ref{thop}), we obtain 
\begin{eqnarray}
&& \frac{\partial f _{ab}}{\partial q^i}
+\epsilon_{ak}\frac{\partial \epsilon_c^{k*}}{\partial {q}^i}
f_{cb} +\epsilon_{bk}^*
 \frac{\partial \epsilon_c^k }{\partial {q}^i} f_{ac}
=  \frac{\partial f _{ab}}{\partial q^i} + 
\frac{2 i e_{\varphi i}}{q \tan\theta} 
\left(\begin{array}{cc} 0 & f_{+-} \\ -f_{-+} & 0
\end{array}\right)_{ab}
\nonumber\\
&& \hspace*{1cm} 
=\,\sum_{l,m} \,(-i)^l \sqrt{\frac{4\pi}{2 l+1}} 
\left\{Y_{lm}^s 
\,\frac{\partial f_{ab,lm}}{\partial q} \,n^i 
+\frac{1}{\sqrt{2}}\frac{f_{ab,lm}}{q}\left(\epsilon^i_{-} \eth_s Y_{lm}^s + 
\epsilon^i_{+} \bar{\eth}_s Y_{lm}^s\right)\right\}.
\label{lmpolterm}
\end{eqnarray}
This form makes explicit that each term carries definite spin, 
such that $s=0$ for the diagonal elements and $s=\pm 2$ for 
the off-diagonals. The derivatives on the spin-weighted spherical
harmonics can be easily taken using (\ref{derYs}).

\section{Expansion of the photon propagation 
term to second order}
\label{sec:expansion}

\noindent
We now turn to the expansion of the Boltzmann equation (\ref{boltzmann}) 
to second order in perturbations. Implementing the simplification of 
the polarization-dependent terms derived in 
Section~\ref{subsec:simplification}, we obtain the first- 
and second-order equations 
\begin{eqnarray}
&& \left[\frac{\partial}{\partial\eta} + 
\frac{q^i}{a E}\frac{\partial}{\partial x^i} \right] 
f_{ab}^{(1)}
+ \left[\frac{1}{P^0} \frac{d q^i}{d\lambda}\right]^{(1)} 
\frac{q^i}{q}\,\frac{\partial f _{ab}^{(0)}}{\partial q} 
= \left[\frac{1}{P^0}\,C_{ab}[f]\right]^{(1)},
\label{boltzmannfirstorder} 
\\[0.4cm]
&& 
\left[\frac{\partial}{\partial\eta} + 
\frac{q^i}{a E}\frac{\partial}{\partial x^i} \right]
f_{ab}^{(2)} 
+ \left[\frac{P^i}{P^0}\right]^{(1)} 
\frac{\partial f_{ab}^{(1)} }{\partial x^i}
+\left[\frac{1}{P^0} \frac{d q^i}{d\lambda} \right]^{(1)}
\left(\frac{\partial f _{ab}^{(1)}}{\partial q^i} +
\epsilon_{ak}\frac{\partial \epsilon_c^{k*}}{\partial {q}^i}
f_{cb}^{(1)} +\epsilon_{bk}^*
 \frac{\partial \epsilon_c^k }{\partial {q}^i} f_{ac}^{(1)}
\right)
\nonumber \\
&&\hspace*{2cm}
+\left[\frac{1}{P^0} \frac{d q^i}{d\lambda} \right]^{(2)}
\frac{q^i}{q}\,\frac{\partial f _{ab}^{(0)}}{\partial q} 
= \left[\frac{1}{P^0}\,C_{ab}[f]\right]^{(2)}.
\label{boltzmannsecondorder} 
\end{eqnarray} 
Here we used that 
\begin{equation}
\frac{1}{P^0}\frac{d x^i}{d\lambda} = \frac{P^i}{P^0} 
= \frac{q^i}{a E} 
\end{equation}
at zeroth order in the perturbation expansion. In this section we 
keep the energy and momentum distinct so that the results also 
apply to the propagation of massive particles.  For photons 
we may use $E=|\bm{p}|=|\bm{q}|/a=q/a$ and $q^i/q=n^i$ to simplify 
the equations.
Eq.~(\ref{boltzmannfirstorder}) reproduces the Boltzmann equation 
in the linear approximation with the familiar free-streaming term 
on the left-hand side. The Fourier transformation converts 
$\partial/\partial x^i \to i k^i$ in the free-streaming terms.
However, the second-order equation also contains products of two 
Fourier-transformed functions, which are to be interpreted as 
convolutions according to (\ref{convolution}). Thus, for instance, 
\begin{equation}
\left[\frac{P^i}{P^0}\right]^{(1)} 
\frac{\partial f_{ab}^{(1)} }{\partial x^i} \to 
\left[\frac{P^i}{P^0}\right]^{(1)}\!\!\!\!\!\!(\bm{k}_1) 
\,i k_2^i \,f_{ab}^{(1)}(\bm{k}_2)
= \int \frac{d^3\bm{k}^\prime}{(2\pi)^3} 
\left[\frac{P^i}{P^0}\right]^{(1)}\!\!\!\!\!\!(\bm{k}-\bm{k}^\prime) 
\,i k^{\prime \,i} f_{ab}^{(1)}(\bm{k}^\prime),
\end{equation} 
in the Fourier transform of (\ref{boltzmannsecondorder}).
In this section we work out the multipole 
transformation of the left-hand side of (\ref{boltzmannsecondorder}). 
The more complicated transformation of the collision term 
is derived in Section~\ref{sec:expcollision}.

\subsection{Covariant momentum and momentum 
derivative}

\noindent
The expression of the covariant momentum in terms of the comoving 
momentum required to evaluate (\ref{boltzmannsecondorder}) is 
obtained from $P^\mu = [e_A]^\mu p^A$. Under the assumptions made in 
this paper (no first-order vector and tensor perturbations, 
conformal Newtonian gauge, observer frame, 
see Section~\ref{subsec:conventions}), we find 
\begin{eqnarray}
P^0 &=& \frac{E}{a}\left(1-A+\frac{3 A^2}{2} - 
\frac{q^i B_i}{a E} + \ldots \right),
\label{p0}
\\
P^i &=& \frac{q^i}{a^2}\left( 1 - D + \frac{3 D^2}{2}  \right) - 
\frac{q^k}{a^2} \,E_{ki} +\ldots,
\label{pi}
\end{eqnarray}
where the ellipses denote corrections of the third-order in 
perturbations. Hence, 
\begin{equation}
\left[\frac{P^i}{P^0}\right]^{(1)} = \frac{q^i}{a E} 
\left(A^{(1)}-D^{(1)}\right).
\label{pip01}
\end{equation}
The change of comoving momentum $dq^i/d\lambda$ along the particle  
trajectory follows from the geodesic equation. We have
\begin{equation}
\frac{dp^i}{d\lambda} = \frac{d([e^i]_\mu P^\mu)}{d\lambda}
= \frac{\partial [e^i]_\mu}{\partial x^\nu} \,P^\nu P^\mu + 
  [e^i]_\mu \left(-\Gamma^\mu_{\nu\rho}P^\nu P^\rho \right) 
= [e^i]_{\mu;\nu} P^\mu P^\nu.
\end{equation}
Then 
\begin{equation}
\frac{1}{P^0} \frac{d q^i}{d\lambda} = \frac{d q^i}{d\eta} 
= \frac{da}{d\eta} \,p^i + \frac{a}{P^0} \frac{dp^i}{d\lambda} 
= H_c \,q^i + a \,[e^i]_{\mu;\nu} \frac{P^\mu P^\nu}{P^0},
\end{equation}
where $H_c=a^{-1} da/d\eta$ denotes the conformal Hubble parameter. 
The previous expression vanishes at zeroth order in the perturbations. 
Its perturbation 
expansion can be calculated from (\ref{p0}), (\ref{pi}) and the 
explicit expressions for the inverse tetrad vectors. The first and 
second order terms required for (\ref{boltzmannsecondorder}) read
\begin{eqnarray}
\left[\frac{1}{P^0} \frac{d q^i}{d\lambda} \right]^{(1)} &=& 
-a E \,\partial^i A^{(1)} - q^i \dot{D}^{(1)} + 
\frac{q^j q^k}{a E}\left(\delta_{jk} \,\partial^i D^{(1)} - 
\delta_{ij} \,\partial^k D^{(1)}\right)
\label{momderivative1}\\
\frac{q^i}{q} \left[\frac{1}{P^0} \frac{d q^i}{d\lambda} \right]^{(2)} 
&=&
-\frac{a E}{q} \,q^i \partial^i A^{(2)} - q \,\dot{D}^{(2)} 
- \frac{q^i q^j}{q}\,\dot{E}_{ij}^{(2)} 
+ \frac{a E}{q} \,q^i \,\dot{B}_i^{(2)} 
+ \frac{(a^2 E^2-q^2)}{q a E} \,q^i H_c B_i^{(2)}
\nonumber\\
&& + \, \frac{a E}{q} \,q^i \partial^i A^{(1)} \left(A^{(1)}+D^{(1)}\right)
+ 2 q \,D^{(1)}\dot{D}^{(1)}.
\label{momderivative2}
\end{eqnarray}
The dot denotes a derivative with respect to conformal time and 
$\partial^i = \partial/\partial x^i$.
The term proportional to $H_c$ in the first line of (\ref{momderivative2})
vanishes for photons and massless propagating particles in general. 

\subsection{Multipole transformation and 
spherical basis}
\label{sec:spherical}

\noindent
The general procedure to obtain the multipole decomposition of 
(\ref{boltzmannfirstorder}), (\ref{boltzmannsecondorder}) is as follows. 
First we insert the representation (\ref{lm1}) for the phase-space 
distributions. Then the direction vector $\bm{n}$ and polarization 
vectors are written in terms of spherical harmonics according to 
\begin{eqnarray}
&&n^i = \sum_m \xi^i_m \sqrt{\frac{4\pi}{3}} Y_{1m},
\qquad
\,n^i n^j = \chi_0^{ij} \sqrt{4\pi} \,Y_{00} + 
 \sum_m \chi_{2m}^{ij}\sqrt{\frac{4\pi}{5}} Y_{2m},
\nonumber\\
&&\epsilon_+^i = \sum_m \xi_m^i  \sqrt{\frac{4\pi}{3}} Y_{1m}^{+1},
\qquad
\epsilon_-^i =- \sum_m \xi_m^i  \sqrt{\frac{4\pi}{3}} Y_{1m}^{-1},
\label{sphericaln}
\end{eqnarray}
which defines $\xi_m^i$ (for $m=0, \pm 1$), 
$\chi_0^{ij}=\frac{1}{3}\delta^{ij}$ and the trace-free tensors 
$\chi_{2m}^{ij}$ (for $m=0,\pm 1,\pm 2$). Explicit expressions 
are provided in Appendix~\ref{xichi}. The multiplication of these 
objects with Cartesian vectors and tensors, respectively, 
projects on the components of the corresponding vectors and tensors in the 
spherical basis. For vectors $V$ and traceless symmetric tensors $T$ 
we define the components in the spherical basis by 
\begin{eqnarray}
&& V_{[0]} = i V_3,
\qquad \hspace*{1cm}
V_{[\pm 1]} = \mp\frac{i}{\sqrt{2}} (V_1\mp i V_2) 
\nonumber\\
&& T_{[0]} = -\frac{3}{2} T_{33},
\hspace*{1.2cm}
T_{[\pm 1]} = \pm \sqrt{2} (T_{13}\mp i T_{23})
\nonumber\\[0.1cm]
&& 
T_{[\pm 2]} = -\frac{1}{\sqrt{6}} (T_{11}-T_{22}\mp 2 i T_{12}).
\end{eqnarray}
Then 
\begin{eqnarray}
&&\xi_m^i V_i = (-i) \,V_{[m]}
\nonumber\\[0.2cm]
&&\chi_{2m}^{ij} T_{ij} = -\alpha_m T_{[m]} \quad \mbox{(no sum over 
$m$)}
\label{sphericalVT}
\end{eqnarray}
with $\alpha_0=\frac{2}{3}$, $\alpha_{\pm 1}=\frac{1}{\sqrt{3}}$ and 
$\alpha_2=1$. At this point, we can use the product formula 
for the spin-weighted spherical harmonics (\ref{eq:Xcgrelation}) to 
express any term in terms of a sum of single harmonics. The 
result of these manipulations is integrated with 
\begin{equation}
L\equiv i^l \sqrt{\frac{2 l+1}{4\pi}} \int d\Omega \,Y_{lm}^{s*}(\bm{n}) 
\label{mL}
\end{equation}
which projects (\ref{boltzmannfirstorder}), (\ref{boltzmannsecondorder}) 
on the $lm$ multipole component. The final step consists of 
transforming from the $ab$ helicity polarization basis to the 
$X=I,V,E,B$ components of the phase-space distribution matrix.

\subsection{Free-streaming term}
\label{sec:freestreaming}

\noindent 
We first consider the three space-time derivative terms in 
(\ref{boltzmannsecondorder}), which after Fourier transformation 
read
\begin{equation}
\frac{\partial f_{ab}^{(2)}}{\partial\eta}, 
\quad
\frac{i \bm{q}\cdot \bm{k}}{a E}\,f_{ab}^{(2)}, 
\quad \left[\frac{P^i}{P^0}\right]^{(1)}\!\!\!\!\!\!(\bm{k}_1) 
\,i k_2^i \,f_{ab}^{(1)}(\bm{k}_2).
\label{freestreaminglist}
\end{equation}
The multipole transformation of the time derivative is trivial since 
\begin{equation}
L\left[\frac{\partial f_{ab}^{(2)}}{\partial\eta}\right] = 
\frac{\partial}{\partial\eta} \,f_{ab,lm}^{(2)}(\bm{k}).
\label{lhsterm1}
\end{equation}
For the transformation of the second term we follow the procedure described 
in Subsection~\ref{sec:spherical}. The manipulations are the same as for 
the corresponding term in the first-order equation 
(\ref{boltzmannfirstorder}), and we discuss them here only to illustrate 
the general method.

Inserting the expansion of $q^i = q n^i$ and $f_{ab}^{(2)}$ 
in spherical harmonics 
gives 
\begin{eqnarray}
\frac{i \bm{q}\cdot \bm{k}}{a E}\,f_{ab}^{(2)} &=& 
\frac{i q}{a E}\,\sum_{m_2} \xi_{m_2}^i  \sqrt{\frac{4\pi}{3}} Y_{1m_2} k^i 
\sum_{l_1,m_1} (-i)^{l_1}\sqrt{\frac{4\pi}{2 l_1+1}} 
\,f_{ab,l_1m_1}^{(2)}(\bm{k}) \,Y^s_{l_1 m_1}
\nonumber\\
&=& \sum_{m_2=-1}^1 \frac{q k^{[m_2]}}{a E}\,
\sum_{l_1,m_1}  (-i)^{l_1}\sqrt{\frac{4\pi}{2 l_1+1}} 
\nonumber\\
&&\times\, 
\sum_{L=|l_1-1|}^{l_1+1} \sum_{S,M=-L}^L 
\frac{\sqrt{2l_1+1}}{\sqrt{2L+1}} 
\,\left(\begin{array}{rcc} l_1 & 1 & L \\ -s & 0 & -S \end{array}\right) 
\left(\begin{array}{ccc} l_1 & 1 & L \\ m_1 & m_2 & M\end{array}\right) 
Y_{LM}^S
\,f_{ab,l_1m_1}^{(2)}(\bm{k}).
\qquad
\end{eqnarray}
Applying the multipole transformation operator $L$ from (\ref{mL}) 
to this expression sets $L=l$ and $M=m$. Interchanging orders of 
summations according to 
\begin{equation}
\sum_{l_1=0}^\infty \sum_{m_1=-l_1}^{l_1} 
\sum_{L=|l_1-1|}^{l_1+1} \sum_{M=-L}^L 
= 
\sum_{L=0}^\infty \sum_{l_1=|L-1|}^{L+1} 
\sum_{M=-L}^L \sum_{m_1=-l_1}^{l_1} 
\end{equation}
yields the final result 
\begin{equation}
L\left[\frac{i \bm{q}\cdot \bm{k}}{a E}\,f_{ab}^{(2)}\right] =
 \sum_{m_2=-1}^1 \frac{q k^{[m_2]}}{a E}\, 
\sum_{l_1=|l-1|}^{l+1} \sum_{m_1=-l_1}^{l_1} i^{l-l_1} 
\,\left(\begin{array}{rcr} l_1 & 1 & l \\ -s & 0 & -s \end{array}\right) 
\left(\begin{array}{ccc} l_1 & 1 & l \\ m_1 & m_2 & m\end{array}\right) 
\,f_{ab,l_1m_1}^{(2)}(\bm{k}).
\label{lhsterm2}
\end{equation}
Here we used that the first Clebsch-Gordan coefficient sets 
$S=s$ to eliminate the sum over $S$. The second one implies $m_2=m-m_1$. 
Recall that $s$ takes the value 0 when $ab=++,--$ and 
$s=\pm 2$ for $ab=\pm\mp$. Thus, in the first case only $l_1=l\pm 1$ 
contribute to the sum, while for the off-diagonal terms $l_1=l$ is 
also non-zero. Eq.~(\ref{lhsterm2}) reproduces the standard first-order 
free-streaming term, in which one usually aligns $\bm{k}$ with the 
three-direction implying $k^{[\pm 1]}=0$ and $k^{[0]}=i k$, 
which simplifies the expression.

The free-streaming term is diagonal in the circular polarization basis, 
but the equations for the two off-diagonal components are slightly 
different, which leads to a mixing of $E$ and $B$ polarization in the 
Stokes parameter basis. The difference arises from 
\begin{equation}
\left(\begin{array}{ccc} l_1 & 1 & l \\ 2 & 0 & 2 \end{array}\right)
= (-1)^{l_1+1-l}
\left(\begin{array}{rrr} l_1 & 1 & l \\ -2 & \phantom{-}0 & -2 
\end{array}\right),
\label{cbg1}
\end{equation}
i.e. when $l_1+1-l$ is odd, which happens precisely for the terms 
with $l_1=l$ present only for $s=\pm 2$. To express the equations 
in the $IVEB$ basis in a compact form we introduce the matrices 
$H_{XY}(l)$ with 
\begin{equation}
H_{XY}(l)=\delta_{XY}\quad \mbox{(for $l$ even)},
\qquad\qquad
H_{XY}(l) = \left(\begin{array}{rrrr}
1&\phantom{-}0&0&\phantom{-}0\\
0&1&0&0\\
0&0&0&i\\
0&0&-i&0
\end{array}\right)\quad \mbox{(for $l$ odd)}
\end{equation}
and define 
\begin{equation}
F_X = \left\{\begin{array}{rc}
0 & \mbox{\qquad $X=I,V$}\\[0.1cm]
-2 &  \mbox{\qquad $X=E,B$}
\end{array}\right.
\end{equation}
Taking linear combinations of (\ref{lhsterm2}) according to 
(\ref{IVEBtransform}) we obtain in the $IVEB$ basis
\begin{eqnarray}
L\left[\frac{i \bm{q}\cdot \bm{k}}{a E}\,f_{X}^{(2)}\right] &=&
 \sum_{m_2=-1}^1 \frac{q k^{[m_2]}}{a E}\, 
\sum_{l_1=|l-1|}^{l+1} \sum_{m_1=-l_1}^{l_1} 
\!i^{l-l_1}
\left(\begin{array}{ccc} l_1 & 1 & l \\ F_X & 0 & F_X \end{array}\right) 
\left(\begin{array}{ccc} l_1 & 1 & l \\ m_1 & m_2 & m\end{array}\right) 
\nonumber\\
&&\times\,
\sum_Y H^*_{XY}(l_1+1-l)\,f_{Y,l_1m_1}^{(2)}(\bm{k}).
\label{lhsterm2IVEB}
\end{eqnarray}
The sum over $Y$ encodes the mixing between the $E$- and $B$-mode 
polarization. 
Since the $H$ matrices are block-diagonal in $IV$ and $EB$, and equal to 
the identity matrix in the $IV$ sector, the sum is redundant for $X=I,V$. 
The equations are decoupled and identical for $X=I$ and $X=V$. 
Nevertheless, the notation introduced above is convenient in order to 
present the results in the $IVEB$ basis without having to resort multiple 
equations for the different cases.

The third term in the list (\ref{freestreaminglist}) requires no 
further work, since using (\ref{pip01})
\begin{equation}
L\left[ \left[\frac{P^i}{P^0}\right]^{(1)}\!\!\!\!\!\!(\bm{k}_1) 
\,i k_2^i \,f_{ab}^{(1)}(\bm{k}_2)\right] = 
\left(A^{(1)}-D^{(1)}\right)\!(\bm{k}_1)\,
L\left[\frac{i \bm{q}\cdot \bm{k}_2}{a E}\,f_{ab}^{(1)}\right].
\label{lhsterm3}
\end{equation}
A convolution of the two mode momenta in the sense of 
(\ref{convolution}) is implied. The application of the $L$-operator 
gives as final result the expression (\ref{lhsterm2}) 
with $k^{[m_2]}\to k_2^{[m_2]}$ and 
$f_{ab,l_1m_1}^{(2)}(\bm{k}) \to f_{ab,l_1m_1}^{(1)}(\bm{k}_2)$, 
or the corresponding result (\ref{lhsterm2IVEB}) in the $IVEB$ basis.  

The generation of $B$ polarization from $E$ polarization through 
free-streaming requires propagation through an inhomogeneous 
universe, and is thus a second-order effect, known as 
time-delay induced $B$ polarization \cite{Hu:2001yq}. 
The time-delay effect is contained 
in the above equations through the off-diagonal terms $H_{BE}^*(\pm 1)=-i$.
The relevant product of Clebsch-Gordan coefficients is 
\begin{equation}
\left(\begin{array}{ccc} l & 1 & l \\ 2 & 0 & 2 \end{array}\right) 
\left(\begin{array}{ccc} l & 1 & l \\ m_1 & m_2 & m\end{array}\right) 
=\delta_{m_1,m-m_2}\times \frac{1}{l (l+1)}\times \left\{\begin{array}{ll}
\,\,2m  & \mbox{\qquad $m_2=0$}\\[0.2cm]
\,\,\mp \sqrt{2 (l+1\mp m)(l\pm m)} &  \mbox{\qquad $m_2=\pm 1$}
\end{array}\right.
\label{cbgrel1}
\end{equation}
In first order in perturbation theory we can always align the mode 
vector $\bm{k}$ such that only $m_2=0$ contributes. 
Then, using (\ref{cbgrel1}) in (\ref{lhsterm2IVEB}), (\ref{lhsterm3})
shows that $EB$ mixing 
occurs only when $m\not =0$, which implies the well-known result 
that no $B$ polarization is induced, when there are no vector or 
tensor perturbations. At second order (\ref{lhsterm3}) contains a 
convolution over all wave-vectors, and the sum over $m_2$ always 
extends over $m_2=0,\pm 1$. It follows from (\ref{cbgrel1}) that 
$EB$ mixing occurs through free-streaming, when the first-order 
scalar perturbations $A^{(1)}$ or $D^{(1)}$ do not vanish. 

To summarize the result of this subsection: the second-order 
space-time derivative terms (free-streaming terms) in the Boltzmann 
equation, Fourier- and multipole-transformed, are given in the 
$IVEB$ basis by
\begin{eqnarray}
&& \frac{\partial}{\partial\eta} \,f_{X,lm}^{(2)}(\bm{k}) \,+ 
 \sum_{m_2=-1}^1  \,
\sum_{l_1=|l-1|}^{l+1} \sum_{m_1=-l_1}^{l_1} 
\!i^{l-l_1}
\left(\begin{array}{ccc} l_1 & 1 & l \\ F_X & 0 & F_X \end{array}\right) 
\left(\begin{array}{ccc} l_1 & 1 & l \\ m_1 & m_2 & m\end{array}\right) 
\nonumber\\
&&\times\,
\sum_Y H^*_{XY}(l_1+1-l)\,
\left[\frac{q k^{[m_2]}}{a E}\,f_{Y,l_1m_1}^{(2)}(\bm{k})
+ \left(A^{(1)}-D^{(1)}\right)\!(\bm{k}_1)\, 
\frac{q k_2^{[m_2]}}{a E}\,f_{Y,l_1m_1}^{(1)}(\bm{k}_2)\right].
\label{lhsterm1to3}
\end{eqnarray}

\subsection{Momentum-derivative terms}

\noindent
We now turn to the multipole decomposition of the two terms involving 
$dq^i/d\lambda$ in (\ref{boltzmannsecondorder}). With the help of 
(\ref{momderivative1}) and (\ref{lmpolterm}) the first one can 
be written as
\begin{eqnarray} 
&&\left[\frac{1}{P^0} \frac{d q^i}{d\lambda} \right]^{(1)}
\left(\frac{\partial f _{ab}^{(1)}}{\partial q^i} +
\epsilon_{ak}\frac{\partial \epsilon_c^{k*}}{\partial {q}^i}
f_{cb}^{(1)} +\epsilon_{bk}^*
 \frac{\partial \epsilon_c^k }{\partial {q}^i} f_{ac}^{(1)}
\right)
\nonumber\\
&& = \, (-1) \sum_{l_1=0}^\infty \sum_{m_1=-l_1}^{l_1} (-i)^{l_1}
\sqrt{\frac{4\pi}{2 l_1+1}} \,\Bigg\{ 
\left[\frac{aE}{q} i\bm{n}\cdot\bm{k}_1 A^{(1)}(\bm{k}_1)+
\dot{D}^{(1)}(\bm{k}_1)\right] Y^s_{l_1 m_1} 
\,q\frac{\partial}{\partial q} f^{(1)}_{ab,l_1 m_1}(\bm{k}_2)
\nonumber\\
&& \hspace*{0.5cm} + \, \left[\frac{aE}{q} i k^i_1 A^{(1)}(\bm{k}_1) - 
\frac{q}{aE}  i k^i_1 D^{(1)}(\bm{k}_1)\right]
\frac{1}{\sqrt{2}}\left(\epsilon^i_{-} \eth_s Y_{l_1 m_1}^s + 
\epsilon^i_{+} \bar{\eth}_s Y_{l_1 m_1}^s\right)  
f^{(1)}_{ab,l_1 m_1}(\bm{k}_2)\Bigg\},
\label{lhsterm4before}
\end{eqnarray}
where we used that $\bm{\epsilon}_\pm$ is orthogonal to $\bm{q}$, and 
$\bm{n}\cdot\bm{q} = q \,\bm{n}^2=q$. Next we express $\bm{n}$ 
and the polarization vectors in terms of spherical harmonics 
according to (\ref{sphericaln}) to write 
\begin{eqnarray}
&& i \bm{n}\cdot\bm{k}_1 = \sum_{m_2=-1}^1 k_1^{[m_2]} 
\sqrt{\frac{4\pi}{3}} Y_{1 m_2} 
\nonumber\\
&& i k^i_1 \left(\epsilon^i_{-} \eth_s Y_{l_1 m_1}^s + 
\epsilon^i_{+} \bar{\eth}_s Y_{l_1 m_1}^s\right) = 
- \!\sum_{m_2=-1}^1 k_1^{[m_2]} 
\sqrt{\frac{4\pi}{3}} \,\Big(
[l_1]_s^+ \,Y_{1 m_2}^{-1} Y^{s+1}_{l_1 m_1}
+[l_1]_s^- \,Y_{1 m_2}^{+1} Y^{s-1}_{l_1 m_1}
\Big)
\end{eqnarray}
after taking the derivatives on the spin-weighted spherical harmonics 
using (\ref{derYs}) in the second equation. 
The remaining steps are straightforward. We eliminate 
the products of spherical harmonics with (\ref{eq:Xcgrelation}) 
and apply the multipole transformation operator (\ref{mL}) to obtain 
\begin{eqnarray}
&& L\Big[\mbox{lhs of (\ref{lhsterm4before})}\Big]_{ab} = 
- \dot{D}^{(1)}(\bm{k}_1)\,
q\frac{\partial}{\partial q} f^{(1)}_{ab,lm}(\bm{k}_2) 
 +\!\sum_{m_2=-1}^1  \,
\sum_{l_1=|l-1|}^{l+1} \sum_{m_1=-l_1}^{l_1} 
\!i^{l-l_1}
\left(\begin{array}{ccc} l_1 & 1 & l \\ m_1 & m_2 & m\end{array}\right) 
\nonumber\\
&&\hspace*{1cm}\times \,\Bigg\{
\left(\begin{array}{rcr} l_1 & 1 & l \\ -s & 0 & -s \end{array}\right) 
\left(-\frac{a E}{q}\right)  k_1^{[m_2]} A^{(1)}(\bm{k}_1)\,
q\frac{\partial}{\partial q} f^{(1)}_{ab,l_1 m_1}(\bm{k}_2)  
\nonumber\\[0.2cm]
&&\hspace*{1.6cm} +\,
\frac{1}{\sqrt{2}}\,\left\{
[l_1]_s^+
\left(\begin{array}{ccr} l_1 & 1 & l \\ -(s+1) & 1 & -s \end{array}\right) 
+
[l_1]_s^-
\left(\begin{array}{crr} l_1 & 1 & l \\ -(s-1) & -1 & -s \end{array}\right) 
\right\}
\nonumber\\[0.2cm]
&&\hspace*{1.6cm} \times\,
 k_1^{[m_2]} \left[\frac{a E}{q} A^{(1)}(\bm{k}_1)-
\frac{q}{aE}  D^{(1)}(\bm{k}_1)\right] f^{(1)}_{ab,l_1 m_1}(\bm{k}_2) 
\Bigg\}.
\label{lhsterm4}
\end{eqnarray}
As for the free-streaming terms the equations for the off-diagonal terms 
are slightly different, which implies conversion of $E$ into $B$ polarization 
and vice versa. The last two lines in the previous equation, which
originate from the derivative of the first-order photon perturbation 
with respect to the direction of the photon momentum, correspond 
precisely to the weak-lensing effect \cite{Lewis:2006fu}.  
If $\kappa(s)$ denotes the expression in curly brackets 
in the third line of (\ref{lhsterm4}), the relation $\kappa(-2) 
=(-1)^{l_1+1-l} \,\kappa(2)$ holds, and because of the similarity 
with (\ref{cbg1}) the same matrix $H_{XY}$ appears in the transformation 
to the Stokes parameters. The final result for this term in the 
$IVEB$ basis reads
\begin{eqnarray}
&& L\Big[\mbox{lhs of (\ref{lhsterm4before})}\Big]_X = 
- \dot{D}^{(1)}(\bm{k}_1)\,
q\frac{\partial}{\partial q} f^{(1)}_{X,lm}(\bm{k}_2) 
 +\!\sum_{m_2=-1}^1  \,
\sum_{l_1=|l-1|}^{l+1} \sum_{m_1=-l_1}^{l_1} 
\!i^{l-l_1}
\left(\begin{array}{ccc} l_1 & 1 & l \\ m_1 & m_2 & m\end{array}\right) 
\nonumber\\
&&\hspace*{1cm}\times \,\Bigg\{
\left(\begin{array}{ccc} l_1 & 1 & l \\ F_X & 0 & F_X \end{array}\right) 
\left(-\frac{a E}{q}\right)  k_1^{[m_2]} A^{(1)}(\bm{k}_1)\,
\sum_Y H^*_{XY}(l_1+1-l)\, 
q\frac{\partial}{\partial q} f^{(1)}_{Y,l_1 m_1}(\bm{k}_2)  
\nonumber\\[0.2cm]
&&\hspace*{1.6cm} +\,
\frac{1}{\sqrt{2}}\,\left\{
[l_1]_{F_X}^-
\left(\begin{array}{ccc} l_1 & 1 & l \\ F_X-1 & 1 & F_X \end{array}\right) 
+
[l_1]_{F_X}^+
\left(\begin{array}{crc} l_1 & 1 & l \\ F_X+1 & -1 & F_X \end{array}\right) 
\right\}
\nonumber\\[0.2cm]
&&\hspace*{1.6cm} \times\,
 k_1^{[m_2]} \left[\frac{a E}{q} A^{(1)}(\bm{k}_1)-
\frac{q}{aE}  D^{(1)}(\bm{k}_1)\right] \sum_Y H^*_{XY}(l_1+1-l)\,
f^{(1)}_{Y,l_1 m_1}(\bm{k}_2) 
\Bigg\}.
\label{lhsterm4IVEB}
\end{eqnarray}

The other momentum-derivative term at second-order can be written as 
\begin{equation}
\left[\frac{1}{P^0} \frac{d q^i}{d\lambda} \right]^{(2)}
\frac{q^i}{q}\,\frac{\partial f _{ab}^{(0)}}{\partial q} 
= \left[X+Y_i\frac{q^i}{q} - \dot{E}^{(2)}_{ij} \,\frac{q^i q^j}{q^2}
\right] \delta_{ab} \,q\frac{\partial f^{(0)}_{I}}{\partial q},
\end{equation}
where $X$ and $Y_i$ represent the $q$-independent and linear terms 
in $q^i$ in (\ref{momderivative2}), respectively. We also used that 
the unperturbed photon phase-space distribution is unpolarized. Since 
the only dependence on the direction of $\bm{q}$ in this term arises 
from the factors of $q^i$ in square brackets, it contributes only 
to $l=0,1,2$. In the quadratic term we write
\begin{equation}
\dot{E}^{(2)}_{ij} \,\frac{q^i q^j}{q^2} = 
\sqrt{\frac{4\pi}{5}} \sum_{m_2=-2}^2 \chi_{2m_2}^{ij} Y_{2 m_2} 
\dot{E}^{(2)}_{ij} = - \sqrt{\frac{4\pi}{5}} \sum_{m_2=-2}^2 
\alpha_{m_2} \dot{E}^{(2)}_{[m_2]}  Y_{2 m_2}, 
\end{equation}
employing the definitions (\ref{sphericaln}), (\ref{sphericalVT}) and 
the tracelessness of $E_{ij}$. The remainder of the calculation is 
straightforward, resulting in 
\begin{eqnarray}
L\left[\left[\frac{1}{P^0} \frac{d q^i}{d\lambda} \right]^{(2)}
\frac{q^i}{q}\,\frac{\partial f _{ab}^{(0)}}{\partial q} \right]
&=& \Bigg\{
\left[-\dot{D}^{(2)}(\bm{k}) + 2 D^{(1)}(\bm{k}_1)\dot{D}^{(1)}(\bm{k}_2) 
\right] \delta_{l0} 
\nonumber\\[0.2cm]
&& \hspace*{-3.5cm} +\, 
\frac{aE}{q} \left[-ik^{[m]} A^{(2)}(\bm{k}) +
i k_1^{[m]} A^{(1)}(\bm{k}_1) \left(A^{(1)}(\bm{k}_2)+D^{(1)}(\bm{k}_2)\right)
+ \dot{B}^{(2)}_{[m]}(\bm{k})+
H_c\left(1-\frac{q^2}{a^2 E^2}\right) B^{(2)}_{[m]}(\bm{k})\right] \delta_{l1} 
\nonumber\\[0.2cm]
&& \hspace*{-3.5cm} 
-\,\alpha_{m} \dot{E}^{(2)}_{[m]}(\bm{k})\,\delta_{l2} 
\Bigg\}  \,\delta_{ab} \,q\frac{\partial f^{(0)}_{I}}{\partial q}.
\label{lhsterm5}
\end{eqnarray}
In the $IVEB$ polarization basis the multipole transform of this term 
takes the same form with the replacement $\delta_{ab}\to \delta_{XI}$.

Our final result for the Boltzmann hierarchy for the multipole moments 
$f^{(2)}_{X,lm}(\bm{k})$ of the Stokes parameter phase-space densities at 
second order is given by the sum of (\ref{lhsterm1to3}), 
(\ref{lhsterm4IVEB}) and (\ref{lhsterm5}) excluding the collision term 
that we consider in the following section. These expressions remain 
valid in the case of massive particles with mass $M$, for which 
\begin{equation}
E = \sqrt{M^2+\frac{q^2}{a^2}}\,
\end{equation}
For photons the simplification $aE/q=1$ can be applied and the term 
proportional to $H_c$ in the second line of (\ref{lhsterm5}) vanishes.

\section{Expansion of the collision term to second order}
\label{sec:expcollision}

\noindent
In this section we compute the expansion of the collision term in the 
Boltzmann hierarchy for the multipole moments 
$f^{(2)}_{X,lm}(\bm{k})$. This is done in two steps. First we expand 
(\ref{scatteringterm}) to second order. Then we apply the operator 
(\ref{mL}) that converts to equations for the multipole 
moments. Our treatment follows \cite{Dodelson:1993xz} extended to 
the polarized phase-space distributions.

\subsection{Non-relativistic expansion}

\noindent
The cosmic background photons that we see have mostly 
last scattered around the 
time of recombination, when the temperature of the universe was less 
than $1\,$eV. Polarization of the CMB is generated at this time or 
later. The electrons on which the photons scatter are therefore 
highly non-relativistic with thermal velocities 
\begin{equation}
\frac{|\bm{q}|}{m_e}\sim \sqrt{\frac{T_e}{m_e}} \approx 10^{-3}.
\label{electronexpansion}
\end{equation}
We therefore perform an expansion of the Compton scattering matrix 
element in the electron momentum and consider the expansion parameter 
(\ref{electronexpansion}) of the same order as the cosmological 
perturbations. Note that in this subsection $\bm{q}$ and $\bm{q}^\prime$ 
refer to the electron momentum and not the comoving photon momentum.

The electrons are in local thermal equilibrium and sufficiently dilute 
to be described by the Maxwell-Boltzmann distribution 
\begin{equation}
g_e(\bm q) = n_e \left(\frac{2\pi}{m_e T_e}\right)^{3/2} 
e^{-\frac{(\bm{q}-m\bm{v}_e)^2}{2 m_e T_e}}.
\label{maxboltz}
\end{equation} 
Here $T_e$, $v_e$ and $n_e$ denote the local electron temperature, 
bulk velocity, and number density of free electrons, i.e. electrons 
not bound in hydrogen or helium. If $x_e$ denotes the ionization 
fraction and $\rho_b$ the baryon density, then $n_e$ is given by 
\begin{equation}
n_e = n_e^{(0)} \left(1+\left[\frac{\delta \rho_b}{\rho_b}\right]^{(1)} + 
\left[\frac{\delta x_e}{x_e}\right]^{(1)} + \ldots\right)
\label{ne}
\end{equation}
to first order in perturbations. A complete account of the collision 
term to second order therefore requires a calculation of the 
recombination history that goes beyond the homogeneous universe to 
obtain the perturbations in the ionization fraction. We refer to 
\cite{Senatore:2008vi} for a discussion of this issue. In our equations we 
keep $n_e$ as an overall factor without expanding it for the time 
being. 

The integral over $\bm{q}^\prime$ in (\ref{scatteringterm}) is 
eliminated by the three-momentum delta-function, which sets 
$\bm{q}^\prime=\bm{q}+\bm{p}-\bm{p}^\prime$. This allows us to 
expand
\begin{eqnarray}
&& g(\bm{q}^\prime\,) = g(\bm{p}+\bm{q}-\bm{p}\,') = 
g(\bm{q}) \left[1-\frac{(\bm{p}-\bm{p}\,')(\bm{q}-m_e\bm{v})}{m_e T}  
-\frac{(\bm{p}-\bm{p}\,')^2}{2m_eT}+
\frac{1}{2}\left(\frac{(\bm{p}-\bm{p}\,')(\bm{q}-m_e\bm{v})}{m_e T}\right)^2
+\ldots \right],\qquad
\nonumber \\[0.2cm]
&& \delta(p^0+q^0-p^{\prime \,0}-q^{\prime \,0}) = 
\delta(p+E(\bm{q})-p'-E(\bm{p}+\bm{q}-\bm{p}\,')) = \delta(p-p') + 
\frac{(\bm{p}-\bm{p}\,')\bm{q}}{m_e}\frac{\partial\delta(p-p')}{\partial p'} 
\nonumber\\ 
&& \hspace*{2cm} +\, \frac{(\bm{p}-\bm{p}\,')^2}{2m_e}
\frac{\partial\delta(p-p')}{\partial p'}
+\frac{1}{2}\left(\frac{(\bm{p}-\bm{p}\,')\bm{q}}{m_e}\right)^2 
\frac{\partial^2 \delta(p-p')}{\partial p'^2}+\ldots,
\label{qexp1}
\end{eqnarray}
where $p=|\bm{p}|$, $p^\prime = |\bm{p}^\prime|$. The expansion is based 
on the observation that $p,p^\prime \sim T$ while $|\bm{q}|\sim 
(m_e T_e)^{1/2}$ and that the difference of electron energies
\begin{equation}
E(\bm{q})-E(\bm{q}\,') = \frac{\bm{q}^2}{2 m_e} - 
\frac{\bm{q}^{\prime \,2}}{2 m_e} = 
- \frac{\bm{q} (\bm{p}-\bm{p}^\prime\,)}{m_e} - 
\frac{(\bm{p}-\bm{p}^\prime\,)^2}{2 m_e} 
\sim \frac{T_e^{3/2}}{m_e} \ll T_e.
\end{equation}
The terms neglected in (\ref{qexp1}) are therefore of third order in the 
expansion parameter (\ref{electronexpansion}). Inserting these expansions 
into the collision term the zeroth-order terms cancel, so that 
the collision term begins at first order as it should be. It is 
therefore sufficient to expand the Compton matrix element to first 
order. The result of expanding (\ref{comptonME}) and (\ref{msquared}) 
can be written in the form 
\begin{equation}
|\overline{M}\,|^2_{\lambda\lambda^\prime;\omega\omega^\prime} = 
24\pi m_e^2\sigma_T\left(S_{0,\lambda\lambda^\prime\omega\omega^\prime} 
+\frac{q^i}{m_e} S^i_{1,\lambda\lambda^\prime\omega\omega^\prime} + 
O\left(\frac{\bm{q}^2}{m_e^2}\right)\right)
\end{equation}
with
\begin{eqnarray}
S_{0,\lambda\lambda^\prime\omega\omega^\prime} &=& 
\bm{\epsilon}_\lambda(\bm{p})\cdot 
\bm{\epsilon}^*_{\lambda^\prime}(\bm{p}^\prime)\,
\bm{\epsilon}^*_\omega(\bm{p})\cdot 
\bm{\epsilon}_{\omega^\prime}(\bm{p}^\prime),
\label{mes0}
\\[0.4cm]
S^i_{1,\lambda\lambda^\prime\omega\omega^\prime} &=&
\bm{\epsilon}_\lambda(\bm{p})\cdot 
\bm{\epsilon}^*_{\lambda^\prime}(\bm{p}^\prime)
\left\{
\epsilon^{*i}_\omega(\bm{p}) \,
\frac{\bm{\epsilon}_{\omega^\prime}(\bm{p}^\prime)\cdot \bm{p}}{p} + 
\epsilon^{i}_{\omega^\prime}(\bm{p}^\prime) \,
\frac{\bm{\epsilon}^*_{\omega}(\bm{p})\cdot \bm{p}^\prime}{p^\prime} 
\right\}
\nonumber\\[0.1cm] 
&& \hspace*{0.0cm} + \, \bm{\epsilon}^*_\omega(\bm{p})\cdot 
\bm{\epsilon}_{\omega^\prime}(\bm{p}^\prime)
\left\{
\epsilon^{i}_\lambda(\bm{p}) \,
\frac{\bm{\epsilon}^*_{\lambda^\prime}(\bm{p}^\prime)\cdot \bm{p}}{p} + 
\epsilon^{*i}_{\lambda^\prime}(\bm{p}^\prime) \,
\frac{\bm{\epsilon}_{\lambda}(\bm{p})\cdot \bm{p}^\prime}{p^\prime} 
\right\},
\label{mes1}
\end{eqnarray}
and $\sigma_T = 8\pi \alpha^2/(3 m_e^2)$ the Thomson scattering cross 
section. At this point the integrand is polynomial in $\bm{q}$ except for 
$g_e(\bm{q})$ so that the integral over $\bm{q}$ in (\ref{scatteringterm}) 
can be expressed in terms of the moments of the electron distribution: 
\begin{equation}
\int\frac{d\bm{q}}{(2\pi)^3} \, g_e(\bm{q})\times 
\big\{1;\,q^i;\,q^i q^j\big\} = 
n_e\times \big\{1;\,m_e v_e^i;\, m_e T_e\delta^{ij}+m_e^2 v_e^i v_e^j\big\}.
\end{equation}

\subsection{Expansion of \boldmath $C_{ab}[f\,]$}

\noindent
It is straightforward to insert the non-relativistic expansions 
discussed above into the collision term (\ref{scatteringterm}) 
and to perform the integrations over the incoming and scattered 
electron momentum. It is convenient to express the results in terms 
of the coefficient of the gain term in (\ref{scatteringterm}) 
and the difference of the gain and loss terms, given by 
\begin{eqnarray}
&& G^{(i)}_{\lambda\lambda^\prime\omega\omega^\prime} =  
f^{(i)}_{\lambda^\prime\omega^\prime}(p^\prime)
\Big[\delta_{a\lambda}(\delta_{\omega b}+f_{\omega b}(p))+
\delta_{\omega b}(\delta_{a\lambda}+f_{a\lambda}(p))\Big],
\nonumber\\[0.2cm]
&& GL^{(i)}_{\lambda\lambda^\prime\omega\omega^\prime} =
2\,\delta_{a\lambda}\delta_{\omega b}\,
f^{(i)}_{\lambda^\prime\omega^\prime}(p^\prime)
- \delta_{\lambda^\prime\omega^\prime }\,
\Big[\delta_{a \lambda} f^{(i)}_{\omega b}(p) +
\delta_{\omega b} f^{(i)}_{a \lambda}(p)\Big]
\label{gainloss}
\end{eqnarray}
at $i$th order in the expansion. Note that while the difference of the gain 
and loss terms is linear in the phase-space distributions, the gain 
term contains quadratic terms. In the definition of 
$G^{(i)}_{\lambda\lambda^\prime\omega\omega^\prime}$ we use the 
unexpanded distribution functions in the Bose enhancement factors. 

The expanded collision term can now be written in the form
\begin{eqnarray}
C_{ab}[f\,] &=&
\frac{3}{4} n_e\sigma_T \int_0^\infty dp^\prime p^\prime 
\int \frac{d\Omega'}{4\pi} \,\Big[c^{(1)}  
+ c_{\Delta}^{(2)}+ c_{v}^{(2)} + c_{\Delta v}^{(2)} 
+ c_{vv}^{(2)} + c_{K}^{(2)}\Big]_{ab},
\label{Cc5}
\end{eqnarray}
where $\Omega^\prime$ denotes the solid angle of the scattered photon 
momentum vector $\bm{p}^\prime$. This expression includes the first-order term
\begin{equation}
c_{ab}^{(1)} = S_{0,\lambda\lambda^\prime\omega\omega^\prime} 
\left[\delta(p-p^\prime)\,GL^{(1)} + 
\bm{v}_e^{(1)}\cdot (\bm{p}-\bm{p}^\prime)\,\frac{\partial\delta(p-p^\prime)}
{\partial p^\prime}\,GL^{(0)}\right]_{
\lambda\lambda^\prime\omega\omega^\prime}
\end{equation}
(summation over repeated photon polarization indices $\lambda,\lambda^\prime,
\omega,\omega^\prime$ is understood), and the second-order term split 
into five contributions according to
\begin{eqnarray}
 c_{\Delta,ab}^{(2)} &=& S_{0,\lambda\lambda^\prime\omega\omega^\prime} 
\,\delta(p-p^\prime)\,GL^{(2)}_{
\lambda\lambda^\prime\omega\omega^\prime}\,
\\[0.4cm]
 c_{v,ab}^{(2)} &=& S_{0,\lambda\lambda^\prime\omega\omega^\prime} 
\,\bm{v}_e^{(2)}\cdot (\bm{p}-\bm{p}^\prime)\,\frac{\partial\delta(p-p^\prime)}
{\partial p^\prime}\,GL^{(0)}_{
\lambda\lambda^\prime\omega\omega^\prime}\,
\\[0.4cm]
 c_{\Delta v,ab}^{(2)} &=& S_{0,\lambda\lambda^\prime\omega\omega^\prime} 
\,\bm{v}_e^{(1)}\cdot (\bm{p}-\bm{p}^\prime)\,\frac{\partial\delta(p-p^\prime)}
{\partial p^\prime}\,GL^{(1)}_{
\lambda\lambda^\prime\omega\omega^\prime}+ 
S^i_{1,\lambda\lambda^\prime\omega\omega^\prime}
\,\delta(p-p^\prime)\,v_e^{(1) i}\,GL^{(1)}_{
\lambda\lambda^\prime\omega\omega^\prime}\, 
\\[0.4cm]
 c_{vv,ab}^{(2)} &=& S_{0,\lambda\lambda^\prime\omega\omega^\prime} 
\,\frac{1}{2}\left[\bm{v}_e^{(1)}\cdot (\bm{p}-\bm{p}^\prime)\right]^2 
\frac{\partial^2 \delta(p-p^\prime)}
{\partial p^{\prime\,2}}\,GL^{(0)}_{
\lambda\lambda^\prime\omega\omega^\prime} 
\nonumber\\ 
&& +\, 
S^i_{1,\lambda\lambda^\prime\omega\omega^\prime} 
\,\bm{v}_e^{(1)}\cdot (\bm{p}-\bm{p}^\prime)\,v_e^{(1) i}\,
\frac{\partial\delta(p-p^\prime)}
{\partial p^\prime}\,GL^{(0)}_{
\lambda\lambda^\prime\omega\omega^\prime}
\\[0.4cm]
 c_{K,ab}^{(2)} &=&  S_{0,\lambda\lambda^\prime\omega\omega^\prime} 
\,\frac{(\bm{p}-\bm{p}^\prime)^2}{2m_e} \left(
\frac{\partial\delta(p-p^\prime)}
{\partial p^\prime}\,GL^{(0)}_{\lambda\lambda^\prime\omega\omega^\prime}-
2 \frac{\partial\delta(p-p^\prime)}
{\partial p^\prime}\,G^{(0)}_{\lambda\lambda^\prime\omega\omega^\prime}+
T_e \frac{\partial^2 \delta(p-p^\prime)}
{\partial p^{\prime\,2}}\,GL^{(0)}_{
\lambda\lambda^\prime\omega\omega^\prime} 
\right)
\nonumber\\
&& +\,
S^i_{1,\lambda\lambda^\prime\omega\omega^\prime} \,
\frac{(p-p^\prime)^i}{m_e}
\left(-\delta(p-p^\prime)\,
G^{(0)}_{\lambda\lambda^\prime\omega\omega^\prime}+
T_e \,\frac{\partial\delta(p-p^\prime)}
{\partial p^\prime}\,GL^{(0)}_{\lambda\lambda^\prime\omega\omega^\prime}
\right).
\label{cK}
\end{eqnarray}
Due to the delta-functions the integral over $p^\prime$ can be performed 
after a few partial integrations. We also sum over polarizations 
and integrate over the solid angle, whenever possible. We define the 
integral operator 
\begin{equation}
I[\ldots] = \frac{1}{2 p} \int_0^\infty dp^\prime p^\prime 
\int \frac{d\Omega'}{4\pi} \,[\ldots]
\end{equation}
such that 
\begin{equation}
C_{ab}[f\,] =
\frac{3}{2} n_e\sigma_T \,p\,\times I\big[c^{(1)}  
+ c_{\Delta}^{(2)}+ c_{v}^{(2)} + c_{\Delta v}^{(2)} 
+ c_{vv}^{(2)} + c_{K}^{(2)}\big]_{ab},
\label{cI}
\end{equation}
and work out the six terms separately.

The first-order term yields 
\begin{equation}
I[c^{(1)}_{ab}] = \frac{1}{2} \int \frac{d\Omega'}{4\pi} \,
 S_{0,\lambda\lambda^\prime\omega\omega^\prime} 
\left[GL^{(1)}_{|p=p^\prime} -
\bm{v}_e^{(1)}\cdot (\bm{n}-2 \bm{n}^\prime)\,GL^{(0)}_{|p=p^\prime}
-\bm{v}_e^{(1)}\cdot (\bm{n}-\bm{n}^\prime) 
\,\,p \frac{\partial}
{\partial p^\prime}\,GL^{(0)}_{|p=p^\prime}\right]_{
\lambda\lambda^\prime\omega\omega^\prime}
\label{fo1}
\end{equation}
after partial integration. We note that $S_0$ and $S_1^i$ depend on the 
direction of $\bm{p}^\prime$ but not on its magnitude $p^\prime$. 
The integral over the delta-function sets 
$\bm{p}^\prime$ to $p \bm{n}^\prime$. The subscript ``$p=p^\prime\,$'' 
means that $f^{(i)}_{ab}(p^\prime) =f^{(i)}_{ab}(p \bm{n}^\prime)$ in the 
expressions (\ref{gainloss}) for the gain and loss terms. The 
zeroth-order distribution function does not depend on the momentum 
direction and is unpolarized, hence 
\begin{equation}
\left[GL^{(0)}_{|p=p^\prime}\right]_{
\lambda\lambda^\prime\omega\omega^\prime} \!\!= \,0,
\qquad\quad 
\left[p \frac{\partial}
{\partial p^\prime}\,GL^{(0)}_{|p=p^\prime}\right]_{
\lambda\lambda^\prime\omega\omega^\prime} \!\!=\, 
2 \,\delta_{a\lambda}\delta_{\omega b}\delta_{\lambda^\prime\omega^\prime}\,
p \frac{\partial f^{(0)}_I}{\partial p}.
\end{equation}
Inserting the expression (\ref{mes0}) for 
$S_{0,\lambda\lambda^\prime\omega\omega^\prime} $ into (\ref{fo1}) 
we next obtain 
\begin{eqnarray}
I[c^{(1)}_{ab}] &=& \frac{1}{2} \,\delta^{ik}\delta^{jl} 
\int \frac{d\Omega'}{4\pi} \bigg\{
2 \,\epsilon^i_a(\bm{n})\epsilon^{*j}_b(\bm{n})
\left[\epsilon^{*k}_{\lambda^\prime}(\bm{n}^\prime)
\epsilon^l_{\omega^\prime}(\bm{n}^\prime) 
f^{(1)}_{\lambda^\prime\omega^\prime}(p \bm{n}^\prime)
- \bm{v}_e^{(1)}\cdot (\bm{n}-\bm{n}^\prime)
\,\epsilon^{*k}_{\lambda^\prime}(\bm{n}^\prime)
\epsilon^l_{\lambda^\prime}(\bm{n}^\prime) \,
p \frac{\partial f^{(0)}_I}{\partial p}\right]
\nonumber \\
&& - \,\epsilon^i_\lambda(\bm{n})\epsilon^{*j}_\omega(\bm{n}) 
\,\epsilon^{*k}_{\lambda^\prime}(\bm{n}^\prime)
\epsilon^l_{\lambda^\prime}(\bm{n}^\prime) \,
\Big[\delta_{a \lambda} f^{(1)}_{\omega b}(p) +
\delta_{\omega b} f^{(1)}_{a \lambda}(p)\Big]
\bigg\}
\end{eqnarray}
This can be further simplified using 
\begin{equation} 
\int \frac{d\Omega'}{4\pi}\,\epsilon^{*k}_{\lambda^\prime}(\bm{n}^\prime)
\epsilon^l_{\lambda^\prime}(\bm{n}^\prime) = 
\int \frac{d\Omega'}{4\pi}\,[\delta^{kl}-n^{\prime\,k}n^{\prime\,l}]
= \frac{2}{3}\,\delta^{kl}
\end{equation}
and $\epsilon^i_\lambda(\bm{n})\epsilon^{*i}_\omega(\bm{n}) = 
\delta_{\lambda\omega}$ to obtain the final result
\begin{eqnarray}
I[c^{(1)}_{ab}] &=& -\frac{2}{3}\,f^{(1)}_{ab}(p) 
-\frac{2}{3}\,\delta_{ab}\, \bm{n}\cdot \bm{v}_e^{(1)} 
\,p \frac{\partial f^{(0)}_I(p)}{\partial p}+ 
\int \frac{d\Omega'}{4\pi} \,\epsilon^i_a(\bm{n})\epsilon^{*j}_b(\bm{n})
\left[\epsilon^{*i}_{\lambda^\prime}(\bm{n}^\prime)
\epsilon^j_{\omega^\prime}(\bm{n}^\prime) \,
f^{(1)}_{\lambda^\prime\omega^\prime}(p \bm{n}^\prime)\right].
\label{i1}
\end{eqnarray}
This expression is equivalent to the standard result for the 
first-order polarized collision term~\cite{Bond:1984fp,Hu:1997mn}.

The second-order terms can be calculated in a similar 
way without further complications though the algebra gets lengthier, when 
the matrix element $S_1^i$ is involved. We also note that the 
gain term alone, which contains quadratic terms in the 
phase-space distributions, appears only in $c_{K,ab}^{(2)}$ at 
second-order, but the simpler zeroth order expression
\begin{equation}
G^{(0)}_{\lambda\lambda^\prime\omega\omega^\prime} =  
2 \,\delta_{a\lambda}\delta_{\omega b}\delta_{\lambda^\prime\omega^\prime}\,
f_I^{(0)}(p^\prime) \Big[1+f_I^{(0)}(p)\Big]
\end{equation}
is needed there. The result for the integrated second-order terms is 
\begin{eqnarray}
I[c^{(2)}_{\Delta, ab}] &=&-\frac{2}{3}\,f^{(2)}_{ab}(p) + 
\int \frac{d\Omega'}{4\pi} \,\epsilon^i_a(\bm{n})\epsilon^{*j}_b(\bm{n})
\left[\epsilon^{*i}_{\lambda^\prime}(\bm{n}^\prime)
\epsilon^j_{\omega^\prime}(\bm{n}^\prime) \,
f^{(2)}_{\lambda^\prime\omega^\prime}(p \bm{n}^\prime)\right],
\label{i2}\\[0.2cm]
I[c^{(2)}_{v, ab}] &=& -\frac{2}{3}\,\delta_{ab}\, \bm{n}\cdot \bm{v}_e^{(2)} 
\,p \frac{\partial f^{(0)}_I(p)}{\partial p}\,
\\[0.2cm]
I[c^{(2)}_{\Delta v, ab}] &=& \frac{2}{3}\, 
\bm{n}\cdot \bm{v}_e^{(1)}\,f^{(1)}_{ab}(p) +
\int \frac{d\Omega'}{4\pi} \left\{
S_{1,ijkl}^m v_e^{(1)m} - \,\delta^{ik}\delta^{jl} 
\left[\bm{v}_e^{(1)}\cdot (\bm{n}-2 \bm{n}^\prime)
+\bm{v}_e^{(1)}\cdot (\bm{n}-\bm{n}^\prime) 
\,\,p \frac{\partial}{\partial p}\right]
\right\}
\nonumber\\
&&\times\,
\epsilon^i_a(\bm{n})\epsilon^{*j}_b(\bm{n})
\left[\epsilon^{*k}_{\lambda^\prime}(\bm{n}^\prime)
\epsilon^l_{\omega^\prime}(\bm{n}^\prime) \,
f^{(1)}_{\lambda^\prime\omega^\prime}(p \bm{n}^\prime)\right],
\label{i4}
\\[0.2cm]
I[c^{(2)}_{vv, ab}] &=& \delta_{ab}\left[\frac{2}{3}\,
{\bm{v}_e^{(1)}}^2+\frac{2}{3}\,(\bm{n}\cdot\bm{v}_e^{(1)})^2\right]
p \frac{\partial f^{(0)}_I(p)}{\partial p} + 
\delta_{ab}\left[\frac{2}{15}\,
{\bm{v}_e^{(1)}}^2+\frac{1}{3}\,(\bm{n}\cdot\bm{v}_e^{(1)})^2\right]
p^2 \frac{\partial^2 f^{(0)}_I(p)}{\partial p^2}\,,
\nonumber\\
&&-\,\epsilon^i_a(\bm{n})\epsilon^{*j}_b(\bm{n})\,\frac{1}{15}\,
v_{e}^{(1)i}v_{e}^{(1)j}\,p^2 \frac{\partial^2 f^{(0)}_I}{\partial p^2}
\\[0.2cm]
I[c^{(2)}_{K, ab}] &=& \frac{2}{3}\delta_{ab}\,\Bigg\{
\frac{4 p}{m_e}\,f_I^{(0)}(p) \big(1+f_I^{(0)}(p)\big)
+\left[\frac{4 T_e}{m_e}+\frac{p}{m_e} \,\big(1+2 f_I^{(0)}(p)\big)\right]
p \frac{\partial f^{(0)}_I(p)}{\partial p} 
\nonumber\\
&& + \,
\frac{T_e}{m_e}\,p^2 \frac{\partial^2 f^{(0)}_I(p)}{\partial p^2}
\Bigg\}.
\label{i6}
\end{eqnarray}
Here $S_{1,ijkl}^m$ equals $S^i_{1,\lambda\lambda^\prime\omega\omega^\prime}$  
with the polarization vectors stripped off, i.e.
\begin{equation}
S_{1,ijkl}^m = \delta^{ik}\left(\delta^{jm} n^l+\delta^{lm}n^{\prime j}
\right)+\delta^{jl}\left(\delta^{im} n^k+\delta^{km}n^{\prime i}
\right).
\end{equation}

\subsection{Fourier- and multipole transformation}

\noindent
The final step in the derivation of the collision term consists in applying 
the multipole transformation operator (\ref{mL}), to perform the 
Fourier-transformation, and to convert the equations to the Stokes 
parameter basis. Taking into account the expansion of the 
prefactors $1/P^0$ in (\ref{boltzmann}) and the free electron density 
$n_e$ in (\ref{Cc5}), the right-hand 
side of the Boltzmann hierachy is of the form
\begin{equation}
\frac{\partial}{\partial\eta} \,f_{X,lm}^{(1)}(\bm{k}) + 
\ldots = L\left[\frac{1}{P^0}\,C_X[f\,]\right] = 
\frac{a}{E}\,L\left[C_X[f\,]^{(1)}\right] 
= \frac{3}{2}\,n_e^{(0)}\sigma_T a \,
\times \hat{I}\big[c^{(1)}_X\big](\bm{k})
\label{rhs1storder}
\end{equation}
at first order, and 
\begin{eqnarray}
\frac{\partial}{\partial\eta} \,f_{X,lm}^{(2)}(\bm{k}) + 
\ldots &=& L\left[\frac{1}{P^0}\,C_X[f\,]\right] = 
\frac{a}{E}\,L\left[C_X[f\,]^{(2)}\right] + 
\frac{a}{E}\,A^{(1)}\,L\left[C_X[f\,]^{(1)}\right]
\nonumber\\[0.2cm]
&& \hspace*{-2cm}\,= \frac{3}{2}\,n_e^{(0)}\sigma_T a \,\times\,\Bigg\{
\hat{I}\left[c_{\Delta}^{(2)}+ c_{v}^{(2)} + c_{\Delta v}^{(2)} 
+ c_{vv}^{(2)} + c_{K}^{(2)}\right]_X\!(\bm{k}) 
\nonumber\\ 
&& \hspace*{-1.5cm}+ \,
\left(A^{(1)}+\left[\frac{\delta \rho_b}{\rho_b}\right]^{(1)} + 
\left[\frac{\delta x_e}{x_e}\right]^{(1)}\,\right) \!(\bm{k}_1)
\,\,\hat{I}\big[c^{(1)}_X\big](\bm{k}_2)
\Bigg\} 
\label{rhs2ndorder}
\end{eqnarray}
at second order. In order to arrive at the last equality in 
each equation we used (\ref{cI}) for the collision term, (\ref{ne}) 
for the free electron density, and set $E=p$, since the collision 
term refers explicitly to photons. We also define the operation 
$\hat I[\ldots]=L[I[\ldots]]$. The Fourier transformation of the 
collision term is trivial, since it does not contain spatial derivatives. 
Products of position-dependent functions simply turn into convolutions 
as indicated by the momentum argument. In the remainder of this
subsection we neglect these arguments to avoid notational
complications, but we restore them in the summary of 
Section~\ref{sec:eqsummary}. To complete the calculation of 
the Boltzmann hierarchy, it now 
remains to work out the multipole transformation 
of (\ref{i1}), and (\ref{i2}) -- (\ref{i6}).

Inserting the multipole expansion (\ref{lm1}) for 
$f^{(1)}_{\lambda^\prime\omega^\prime}(p \bm{n}^\prime)$ and 
applying the $L$-operator (\ref{mL}) the angular integrals  
can be expressed in terms of the matrices
\begin{equation}
Q^{ij}_{ab,lm} = \frac{1}{\sqrt{4\pi}} \int d\Omega \,
\epsilon_b^{*i}(\bm{n})\epsilon_a^j(\bm{n}) \,Y^s_{lm}(\bm{n}),
\label{Qmatrices}
\end{equation}
where $s=0$ is implied for $ab=++,--$ and $s=\pm 2$ for 
$ab=\mp\pm$. Since the polarization vectors are spin-1 objects, 
the $Q$-matrices are non-zero only for $l\leq 2$. To transform 
to the Stokes parameter basis we use the matrices 
(\ref{Umatrices}) and define
\begin{equation}
Q^{ij}_{X,lm} = U_{X,[ab]} Q_{[ab],lm} 
\label{QmatricesIVEB}
\end{equation}
in analogy with (\ref{IVEBtransform}) for the phase-space distributions. 
In the $IVEB$ basis the $Q$-matrices vanish for $X=B$ for any value of $l$.
The non-vanishing $Q$-matrices are given explicitly 
in (\ref{eq:XQsexplizit}). The trace 
\begin{equation}
\mbox{tr}\left(Q^\dagger_{ab,lm} Q_{cd,l^\prime m^\prime}\right) 
=Q^{ij*}_{ab,lm} Q^{ij}_{cd,l^\prime m^\prime} 
\equiv \frac{1}{3}\,\omega^{(l)}_{ab;cd}\,\delta_{l l^\prime}
\delta_{m m^\prime}
\label{wsymbolsab}
\end{equation}
is diagonal in the multipole indices and defines the $\omega$-symbols. 
Similarly in the $IVEB$ basis
\begin{equation}
\mbox{tr}\left(Q^\dagger_{X,lm} Q_{Y,l^\prime m^\prime}\right) 
=Q^{ij*}_{X,lm} Q^{ij}_{Y,l^\prime m^\prime} 
\equiv \frac{1}{3}\,\omega^{(l)}_{XY}\,\delta_{l l^\prime}
\delta_{m m^\prime}
\label{wsymbols}
\end{equation}
for 
\begin{equation}
\omega^{(l)}_{XY} = U_{X;[ab]}^* U_{Y;[cd]} \omega^{(l)}_{ab;cd}.
\end{equation}
Only very few of the $\omega^{(l)}_{XY}$ are not zero. In particular, 
$\omega^{(2)}_{IE}=\omega^{(2)}_{IE}=-\sqrt{3/50}$ is the only 
off-diagonal term that couples to polarization. The other 
non-vanishing values are summarized in (\ref{wsymbolvalues}).

After these preliminaries we turn to the explicit calculation beginning 
with the first-order term. With the definitions (\ref{Qmatrices}) 
and (\ref{wsymbolsab}) it is straightforward to obtain from 
(\ref{i1}) the expression
\begin{equation}
\hat I\big[c^{(1)}_{ab}\big] = 
\frac{2}{3}\left\{
-f^{(1)}_{ab,lm}(p) - \delta_{ab} \delta_{l1} \, v_{e,[m]}^{(1)}
\,p \frac{\partial f^{(0)}_I(p)}{\partial p}+ 
\frac{1}{2}\,\omega^{(l)}_{ba;\omega^\prime\lambda^\prime}
f^{(1)}_{\lambda^\prime\omega^\prime,lm}(p)
\right\}.
\end{equation}
Note that the phase-space distributions $f^{(1)}_{ab,lm}(p)$ and 
helicity components 
of the electron bulk-velocity field $v_{e,[m]}^{(1)}$ also 
depend on the Fourier mode vector $\bm{k}$. This dependence is 
suppressed in this subsection as mentioned above. The transformation to 
the Stokes parameter basis requires the calculation of 
$U_{X,[ab]} \,\hat I\big[c^{(1)}\big]_{ab}$. For this purpose we 
use 
\begin{eqnarray}
&& U_{X,[ab]}\omega^{(l)}_{ba;\omega^\prime\lambda^\prime}
f^{(1)}_{\lambda^\prime\omega^\prime,lm}(p) =
U_{X,[ab]} U^{-1 *}_{[ba],X^\prime} \,\omega^{(l)}_{X^\prime Y^\prime} 
U^{-1}_{[\omega^\prime\lambda^\prime];Y^\prime}
U^{-1}_{[\lambda^\prime\omega^\prime],Y} f^{(1)}_{Y,lm}(p)
\nonumber\\
&& \hspace*{1cm} \stackrel{(\ref{Urel})}{=}\,
2\,U_{X,[ab]} U^{-1}_{[ab],X^\prime} \,\omega^{(l)}_{X^\prime Y^\prime} 
U^{-1 *}_{[\lambda^\prime\omega^\prime];Y^\prime}
U^{*}_{Y,[\lambda^\prime\omega^\prime]} f^{(1)}_{Y,lm}(p)
= 2\,\delta_{XX^\prime} \,\omega^{(l)}_{X^\prime Y^\prime} 
\delta_{Y^\prime Y} f^{(1)}_{Y,lm}(p) 
\nonumber\\
&& \hspace*{1cm} =\,  
2\,\omega^{(l)}_{XY} f^{(1)}_{Y,lm}(p)
\end{eqnarray}
and obtain
\begin{equation}
\hat I\big[c^{(1)}_X\big] = 
\frac{2}{3}\left\{
-f^{(1)}_{X,lm}(p) - \delta_{XI} \,\delta_{l1} \, v_{e,[m]}^{(1)}
\,p \frac{\partial f^{(0)}_I(p)}{\partial p}+ 
\omega^{(l)}_{XY} f^{(1)}_{Y,lm}(p)
\right\}.
\label{ih1}
\end{equation}
When inserted into (\ref{rhs1storder}) we reproduce the 
first-order collision term in the Boltzmann hierarchy for the 
polarized phase-space distributions in a notation similar to 
\cite{seljak}. The last term in brackets 
describes the generation of the $E$-polarization quadrupole in 
Thomson scattering.

The second-order terms can be calculated in a similar way. The first two, 
$I[c^{(2)}_{\Delta, ab}]$ and $I[c^{(2)}_{v, ab}]$, have the same structure 
as the first-order term and can be obtained from (\ref{ih1}) 
without additional work:
\begin{eqnarray}
\hat I[c^{(2)}_{\Delta, X}] &=& \frac{2}{3}\left\{
-f^{(2)}_{X,lm}(p) + 
\omega^{(l)}_{XY} f^{(2)}_{Y,lm}(p)
\right\},
\label{ih2}
\\[0.2cm]
\hat I[c^{(2)}_{v, X}] &=& -\frac{2}{3}\,\delta_{XI} \,\delta_{l1} \, 
v_{e,[m]}^{(2)}\,p \frac{\partial f^{(0)}_I(p)}{\partial p}.
\label{ih3}
\end{eqnarray}
By far the most complicated expression to transform to multipole variables 
is the term $I[c^{(2)}_{\Delta v, ab}]$ in (\ref{i4}). In terms of the 
$Q$-matrices and $\omega$-coefficients introduced before we find 
\begin{eqnarray}
\hat I[c^{(2)}_{\Delta v, X}] &=& \frac{2}{3} 
\sum_{m_2=-1}^1  \,
\sum_{l_1=|l-1|}^{l+1} \sum_{m_1=-l_1}^{l_1} 
\!i^{l-l_1-1}
\,v_{e,[m_2]}^{(1)} 
\nonumber\\
&&\hspace*{0cm}\times \,\Bigg\{
\left(\begin{array}{ccc} l_1 & 1 & l \\ m_1 & m_2 & m\end{array}\right) 
\bigg[\left(\begin{array}{ccc} l_1 & 1 & l \\ F_X & 0 & F_X \end{array}\right) 
\sum_Y H^*_{XY}(l_1+1-l)\,f^{(1)}_{Y,l_1 m_1}(p) 
\nonumber\\
&&\hspace*{0.8cm} +\,
\sum_{Y,Z} 
\left(\begin{array}{ccc} l_1 & 1 & l \\ F_Y & 0 & F_Y \end{array}\right) 
\omega^{(l)}_{XZ}\,H^*_{ZY}(l_1+1-l)
\left(2 f^{(1)}_{Y,l_1 m_1}(p)+ 
p\frac{\partial}{\partial p} f^{(1)}_{Y,l_1 m_1}(p)\right)
\bigg]
\nonumber\\
&&\hspace*{0.8cm} -\,
(-1)^{m_2}\frac{2 l+1}{2 l_1+1} 
\left(\begin{array}{ccc} l & 1 & l_1 \\ m & -m_2 & m_1\end{array}\right) 
\left(\begin{array}{ccc} l & 1 & l_1 \\ F_X & 0 & F_X \end{array}\right)
\nonumber\\
&&\hspace*{1.1cm}\times
\sum_{Y,Z} H^*_{XZ}(l+1-l_1)\,\omega^{(l_1)}_{ZY}
\left(f^{(1)}_{Y,l_1 m_1}(p)+ 
p\frac{\partial}{\partial p} f^{(1)}_{Y,l_1 m_1}(p)\right)\Bigg\}
\nonumber\\
&& +\,\frac{2}{3} 
\sum_{m_2=-1}^1  \,
\sum_{l_1=|l-1|}^{l+1} \sum_{m_1=-l_1}^{l_1} 
\!i^{l-l_1}\,\sum_{Y,Z} 3\,f^{(1)}_{Y,l_1 m_1}(p)
\nonumber\\
&&\hspace*{0cm}\times \,\Bigg\{
(-1)^{m_2}\sum_{L=|l-1|}^{l+1}\sum_{M=-L}^L 
\frac{2 l+1}{\sqrt{(2 l_1+1)(2 L+1)}} 
\left(\begin{array}{ccc} l & 1 & L \\ m & -m_2 & M\end{array}\right) 
\left(\begin{array}{ccc} l & 1 & L \\ F_X & 0 & F_X \end{array}\right)
\nonumber\\
&&\hspace*{0.6cm}\times\,
H^*_{XZ}(l+1-L)
\left[v_e^{(1)j} \xi_{m_2}^k Q_{Z,LM}^{ij*} Q_{Y,l_1 m_1}^{ik} +
v_e^{(1)i} \xi_{m_2}^k Q_{Z,LM}^{ij*} Q_{Y,l_1 m_1}^{kj}\right] 
\nonumber\\
&&\hspace*{0.6cm}
+ \,\sum_{L=|l_1-1|}^{l_1+1}\sum_{M=-L}^L 
\sqrt{\frac{2 l+1}{2 L+1}} 
\left(\begin{array}{ccc} l_1 & 1 & L \\ m_1 & m_2 & M\end{array}\right) 
\left(\begin{array}{ccc} l_1 & 1 & L \\ F_Y & 0 & F_Y \end{array}\right) 
\nonumber\\
&&\hspace*{0.6cm}\times
\left[v_e^{(1)k} \xi_{m_2}^j Q_{X,lm}^{ij*} Q_{Z,LM}^{ik} +
v_e^{(1)k} \xi_{m_2}^i Q_{X,lm}^{ij*} Q_{Z,LM}^{kj}\right] 
H^*_{ZY}(l_1+1-L)\,\Bigg\}.
\label{ih4}
\end{eqnarray}
We have made the sums over $Y$, $Z$ explicit here. 
While providing a closed expression for any $X=I,V,E,B$, this result is 
not very transparent. Recalling that the $Q$-matrices and 
$\omega$-coefficients vanish for $l,l_1,L>2$ and noting that the 
Clebsch-Gordan coefficients are non-zero only if the angular momenta 
differ by no more than one, we see that $\hat I[c^{(2)}_{\Delta v, X}]$ 
vanishes when $l>3$. For any particular $X$ the sums can be worked out 
explicitly at the expense of introducing explicit values of the 
Clebsch-Gordan coefficients. We give the corresponding simpler 
expressions in our summary of the Boltzmann hierachy in 
Section~\ref{sec:eqsummary}.
The last two terms to be converted to the multipole representation, 
$I[c^{(2)}_{vv, ab}]$ and $I[c^{(2)}_{K, ab}]$, are relatively 
simple, since they depend only on the unperturbed phase-space densities. 
In particular, $I[c^{(2)}_{K, ab}]$ does not contain any perturbation 
variables, hence angular dependence, since it arises from the 
non-relativistic expansion. The result is
\begin{eqnarray}
\hat I[c^{(2)}_{vv, X}] &=& \frac{2}{3} \,\Bigg\{
\delta_{XI} \,p \frac{\partial f^{(0)}_I(p)}{\partial p}
\,\bigg[\delta_{l0}\delta_{m0}\,{\bm{v}_e^{(1)}}^2 
- \sum_{m_1,m_2=-1}^1 v_{e,[m_1]}^{(1)}\,v_{e,[m_2]}^{(1)}\,i^l
\left(\begin{array}{ccc} 1 & 1 & l \\ m_1 & m_2 & m\end{array}\right) 
\left(\begin{array}{ccc} 1 & 1 & l \\ 0 & 0 & 0\end{array}\right) 
\bigg]
\nonumber\\
&& \hspace*{0.6cm}+ \,
\delta_{XI} \,p^2 \frac{\partial^2 f^{(0)}_I(p)}{\partial p^2}
\,\bigg[\delta_{l0}\delta_{m0}\,\frac{1}{5}\,{\bm{v}_e^{(1)}}^2 
- \sum_{m_1,m_2=-1}^1 
\frac{1}{2}\,v_{e,[m_1]}^{(1)}\,v_{e,[m_2]}^{(1)}\,i^l
\left(\begin{array}{ccc} 1 & 1 & l \\ m_1 & m_2 & m\end{array}\right) 
\left(\begin{array}{ccc} 1 & 1 & l \\ 0 & 0 & 0\end{array}\right) 
\bigg]
\nonumber\\
&&  \hspace*{0.6cm}- \,i^l\sqrt{2 l+1}\,\frac{1}{10} v_e^{(1)i}v_e^{(1)j} 
\,Q^{ij*}_{X,lm}\,p^2 \frac{\partial^2 f^{(0)}_I(p)}{\partial p^2}
\Bigg\},
\label{ih5}
\\[0.2cm]
\hat I[c^{(2)}_{K, X}] &=& \frac{2}{3}\,\delta_{XI}\,\delta_{l0}\delta_{m0}
\,\Bigg\{\frac{4 p}{m_e}\,f_I^{(0)}(p) \big(1+f_I^{(0)}(p)\big)
+\left[\frac{4 T_e}{m_e}+\frac{p}{m_e} \,\big(1+2 f_I^{(0)}(p)\big)\right]
p \frac{\partial f^{(0)}_I(p)}{\partial p} 
\nonumber\\
&& + \,
\frac{T_e}{m_e}\,p^2 \frac{\partial^2 f^{(0)}_I(p)}{\partial p^2}
\Bigg\}.
\label{ih6}
\end{eqnarray}
Here and above in (\ref{ih4}) 
we expressed the result directly in the Stokes parameter 
basis. The result in the circular polarization basis is obtained by 
omitting the sums over $Y,Z$ and replacing $H^*_{PQ}(\ldots)\to 
\delta_{PQ}$ (any $P,Q$); by 
substituting $\delta_{XI}\to\delta_{ab}$, 
$X\to ab$, $Y\to \lambda^\prime\omega^\prime$,  
 $F_X\to -s$, $F_Y\to -s^\prime$ (with $s$ chosen 
according to the value of $ab$ and $s^\prime$ 
according to $\lambda^\prime\omega^\prime$), 
as well as 
$\omega^{(l)}_{XY}\to \omega^{(l)}_{ba,\omega^\prime\lambda^\prime}/2$, 
$Q^{ij*}_{X,lm} \to Q^{ij*}_{ba,lm}$, 
$Q^{ij}_{Y,lm} \to Q^{ij}_{\omega^\prime\lambda^\prime,lm}$ 
(any $i,j,l,m$) in (\ref{ih1}) -- (\ref{ih6}).

\section{Boltzmann hierachy at second order -- summary of equations}
\label{sec:eqsummary}

\noindent 
At this point we can return to our original notation and express 
the phase-space densities in terms of the comoving momentum $q=a p$, 
using $p \,\partial f_X(p)/\partial p = q \,\partial f_X/\partial q$. 
In the following we leave away the photon momentum argument $q$ on 
the phase-space densities but restore the Fourier mode momentum 
writing $f_{X,lm}(\bm{k}_i)$. As mentioned above by taking the 
four values of $X$ separately, we can evaluate the angular momentum 
sums over $l$, $l^\prime$ etc., and obtain a more explicit form 
of the Boltzmann hierarchy. We summarize the second-order equations in 
this section. For convenience we recall the first-order equations 
in the absence of first-order vector and tensor modes 
in the present notation: 
\begin{eqnarray}
\label{Imode1storder}
\frac{\partial}{\partial \eta}f_{I,lm}^{(1)}(\bm{k}) &+& 
\sum_{\pm} (\mp i) f_{I(l \pm 1)m_1}^{(1)}(\bm{k})
k^{[m_2]}C_{m_1  m}^{\pm,l} 
- \delta_{l0} q\frac{\partial f_I^{(0)}}{\partial q}\dot{D}^{(1)}(\bm{k}) 
- i \delta_{l1}q\frac{\partial f_I^{(0)}}{\partial
  q}k^{[m]}A^{(1)}(\bm{k}) 
\nonumber \\
=|\dot{\kappa}|\,\bigg\{&-& 
f_{I,lm}^{(1)}(\bm{k}) +\delta_{l0}f_{I,00}^{(1)}(\bm{k})  
-\delta_{l1}q\frac{\partial f_I^{(0)} }{\partial q}
v^{(1)}_{e,[m]}(\bm{k}) 
+ \delta_{l2}\frac{1}{10}\left(f_{I,2m}^{(1)}(\bm{k}) 
- \sqrt{6}f_{E,2m}^{(1)}(\bm{k}) \right)\bigg\} 
\\[0.4cm]
\frac{\partial}{\partial \eta}f_{V,lm}^{(1)}(\bm{k}) &+& 
\sum_{\pm} (\mp i) f_{V,(l \pm 1)m_1}^{(1)}(\bm{k})
k^{[m_2]}C_{m_1 m}^{\pm,l}
=|\dot{\kappa}|\,\bigg\{ -f_{V,lm}^{(1)}(\bm{k}) +
\delta_{l1}\frac{1}{2}f^{(1)}_{V,1m}(\bm{k})\bigg\}
\\[0.4cm]
\frac{\partial}{\partial \eta}f_{E,lm}^{(1)}(\bm{k}) &+& 
\sum_{\pm} (\mp i) f_{E,(l \pm 1)m_1}^{(1)}(\bm{k})
k^{[m_2]}D_{m_1 m}^{\pm,l} 
- i f_{B,l m_1}^{(1)}(\bm{k}) k^{[m_2]} D^{0,l}_{m_1 m}  
\nonumber \\
=|\dot{\kappa}|\,\bigg\{ &-& f_{E,lm}^{(1)}(\bm{k}) 
-\delta_{l2} \frac{\sqrt{6}}{10} 
\left(f_{I,2m}^{(1)}(\bm{k}) -
\sqrt{6}f_{E,2m}^{(1)}(\bm{k})\right)\bigg\} 
\label{Emode1storder}
\\[0.4cm]
\frac{\partial}{\partial \eta}f_{B,lm}^{(1)}(\bm{k}) &+& 
\sum_{\pm} (\mp i) f_{B,(l \pm 1)m_1}^{(1)}(\bm{k}) 
k^{[m_2]} D^{\pm,l}_{m_1 m}  
+ i f_{E,l m_1}^{(1)}(\bm{k}) k^{[m_2]}D_{m_1 m}^{0,l} 
=|\dot{\kappa}|\,\bigg\{ - f_{B,lm}^{(1)}(\bm{k})\bigg\}
\label{Bmode1storder}
\end{eqnarray}
Here we introduced the abbreviation
\begin{equation}
\dot{\kappa} = -n_e^{(0)} \sigma_T a <0
\end{equation}
for the collision rate. Furthermore, here and below 
a summation over $m_2=0,\pm 1$ is implicitly understood in terms 
containing the index $m_2$, 
and $m_1$ is equal to $m-m_2$. We also introduce the coupling 
coefficients
\begin{eqnarray}
C^{+,l}_{m\pm 1,m} &=& 
-\frac{\sqrt{(l+1\pm m)(l+2\pm m)}}{\sqrt{2}(2l+3)} 
\nonumber \\
C^{+,l}_{m,m} &=& \frac{\sqrt{(l+1)^2-m^2}}{2l+3} 
\nonumber \\
C^{-,l}_{m\pm 1,m} &=& \frac{\sqrt{(l-1\mp m)(l\mp m)}}{\sqrt{2}(2l-1)} 
\nonumber \\
C^{-,l}_{m,m} &=& \frac{\sqrt{l^2-m^2}}{2l-1} 
\nonumber \\
D^{+,l}_{m_1 m} &=& \frac{\sqrt{(l-1)(l+3)}}{l+1}C^{+,l}_{m_1 m} 
\nonumber \\
D^{-,l}_{m_1 m} &=& \frac{\sqrt{l^2-4}}{l}C^{-,l}_{m_1 m} 
\nonumber \\
D^{0,l}_{m\pm 1,m} &=& \mp \frac{\sqrt{2(l+1\pm m)(l \mp m)}}{l (l+1)} 
\nonumber \\
 D^{0,l}_{m,m} &=& -\frac{2m}{l (l+1)}, 
\end{eqnarray}
as well as 
\begin{eqnarray}
&& R^{+,l}_{m_1 m} = -(l+2) C^{+,l}_{m_1 m},
\qquad\quad  R^{-,l}_{m_1 m} =(l-1) C^{-,l}_{m_1 m},
\nonumber\\[0.2cm]
&& K^{+,l}_{m_1 m} = -(l+2) D^{+,l}_{m_1 m},
\qquad\quad  K^{-,l}_{m_1 m} = (l-1) D^{-,l}_{m_1 m},
\nonumber\\[0.2cm]
&&\hspace*{3cm}  K^{0,l}_{m_1 m} = - D^{0,l}_{m_1 m}.
\qquad
\end{eqnarray}
Note that we may choose $\bm{k}$ such that it points into the 
three-direction, in which case $k_{[\pm 1]}=0$ and the first-order 
equations become particularly simple.

We now present our main result, the Boltzmann hierachy for the
second-order perturbations to the polarized phase-space densities.
Recall that the equations are given in conformal Newtonian gauge 
for a comoving and aligned observer ($U_i=0, \theta_i=0$) under 
the assumptions of vanishing first-order vector and tensor modes 
($B_i^{(1)}=E_{ij}^{(1)}=0$). In the equations given below we keep 
terms involving the first-order perturbations $f_{B,lm}^{(1)}$ of 
the $B$-polarization density to display their structure. Of course, 
under the above assumptions there is no $B$ polarization in 
first order, so $f_{B,lm}^{(1)}$ vanishes, and the 
corresponding terms can be neglected in numerical evaluations. 
The equations read:
\begin{eqnarray}
\frac{\partial}{\partial \eta}f_{I,lm}^{(2)}(\bm{k}) &+& 
\sum_{\pm} \,(\mp i) f_{I,(l \pm 1)m_1}^{(2)}(\bm{k})
\,k^{[m_2]}C_{m_1 m}^{\pm,l} 
- \delta_{l0} \,q\frac{\partial f_I^{(0)}}{\partial q} \dot{D}^{(2)}(\bm{k}) 
\nonumber \\
&+& \delta_{l1}\,q\frac{\partial f_I^{(0)}}{\partial q} 
\left(-i k_{[m]}A^{(2)}(\bm{k}) + \dot{B}_{[m]}^{(2)}(\bm{k})\right)  
- \delta_{l2}\,q\frac{\partial f_I^{(0)}}{\partial q}
\alpha_m\dot{E}^{(2)}_{[m]}(\bm{k})
\nonumber \\
&-&\dot{D}^{(1)}(\bm{k}_1)\,
q\frac{\partial}{\partial q}f_{I,lm}^{(1)}(\bm{k}_2) 
+\sum_{\pm} \,(\mp i) \bigg\{k_1^{[m_2]}\left(A^{(1)}-D^{(1)}
\right)(\bm{k}_1)
f_{I,(l \pm 1)m_1}^{(1)}(\bm{k}_2) R_{m_1 m}^{\pm, l} 
\nonumber \\
&& \hspace*{0cm} 
+\,\bigg(k_2^{[m_2]}\left(A^{(1)}-D^{(1)}\right)(\bm{k}_1)
-k_1^{[m_2]}A^{(1)}(\bm{k}_1)\,q\frac{\partial}{\partial
  q}\bigg)\,f_{I,(l \pm 1)m_1}^{(1)}(\bm{k}_2) C_{m_1 m}^{\pm,l}\bigg\} 
\nonumber \\[0.1cm]
&+& 2 \delta_{l0} \,q\frac{\partial  f_I^{(0)}}{\partial q} 
\dot{D}^{(1)}(\bm{k}_1)D^{(1)}(\bm{k}_2) 
+ \delta_{l1}\,q\frac{\partial f_I^{(0)}}{\partial q}\left( i
  k_1^{[m]}A^{(1)}(\bm{k}_1)\left(A^{(1)}+D^{(1)}
  \right)(\bm{k}_2)\right) 
\nonumber \\
=|\dot{\kappa}|\,\Bigg\{&-& f_{I,lm}^{(2)}(\bm{k}) 
+\delta_{l0}f_{I,00}^{(2)}(\bm{k})  
-\delta_{l1}\,q\frac{\partial f_I^{(0)}}{\partial q} v_{e,[m]}^{(2)}(\bm{k}) 
+ \delta_{l2}\frac{1}{10}\left(f_{I,2m}^{(2)}(\bm{k}) 
- \sqrt{6}f_{E,2m}^{(2)}(\bm{k}) \right)
\nonumber \\
&+& 
\bigg(A^{(1)}+\left[\frac{\delta \rho_b}{\rho_b}\right]^{(1)} + 
\left[\frac{\delta x_e}{x_e}\right]^{(1)}\,\bigg)(\bm{k}_1)
\,\bigg(-f_{I,lm}^{(1)}(\bm{k}_2) 
+\delta_{l0}f_{I,00}^{(1)}(\bm{k}_2)  
-\delta_{l1}\,q\frac{\partial f_I^{(0)}}{\partial q} v_{e,[m]}^{(1)}(\bm{k}_2) 
\nonumber \\[0.1cm]
&&\hspace*{0cm} 
+\, \delta_{l2}\frac{1}{10}\left(f_{I,2m}^{(1)} 
- \sqrt{6}f_{E,2m}^{(1)} \right)(\bm{k}_2)\bigg)
\nonumber \\[0.1cm]
&+&  \sum_{\pm}\,(\mp 1) v^{(1)}_{e,[m_2]}(\bm{k}_1)
f_{I,(l \pm 1)m_1}^{(1)}(\bm{k}_2) C^{\pm,l}_{m_1 m}
\nonumber \\
&+&\delta_{l0}\,\Bigg\{- v^{(1)}_{e,[m_2]}(\bm{k}_1)
\left(q\frac{\partial}{\partial q}f_{I,1m_1}^{(1)}+
2f_{I,1m_1}^{(1)}\right)(\bm{k}_2)C^{+,0}_{m_1 m}
\nonumber \\
&&\hspace*{0.85cm} 
-\,v^{(1)}_{e,[m_1]}(\bm{k}_1)v^{(1)}_{e,[m_2]}(\bm{k}_2)
\left(4q\frac{\partial f_I^{(0)}}{\partial q}+
q^2\frac{\partial^2 f_I^{(0)}}{\partial q^2}
\right)C^{+,0}_{m_1 m} 
\nonumber \\
&&\hspace*{0.85cm} 
+\,\frac{4 q}{m_e}\,f_I^{(0)} \big(1+f_I^{(0)}\big)
+\left[\frac{4 T_e}{m_e}+\frac{q}{m_e} \,\big(1+2 f_I^{(0)}\big)\right]
q \frac{\partial f^{(0)}_I }{\partial q}
+ \frac{T_e}{m_e}\,q^2 \frac{\partial^2 f^{(0)}_I }{\partial q^2}\Bigg\} 
\nonumber \\
&+&\delta_{l1}\, v^{(1)}_{e,[m_2]}(\bm{k}_1)\,\Bigg\{
-\left(q\frac{\partial}{\partial q}+1\right)
f_{I,0m_1}^{(1)}(\bm{k}_2)C^{-,1}_{m_1 m} 
\nonumber \\
&&\hspace*{0.85cm}
+\,\frac{1}{10}
\left(q\frac{\partial}{\partial q}+4\right)\left(f^{(1)}_{I,2m_1}
-\sqrt{6}f^{(1)}_{E,2m_1}\right)(\bm{k}_2)C^{+,1}_{m_1 m} \Bigg\} 
\nonumber \\
&+& \delta_{l2} \,\Bigg\{\frac{1}{10}v^{(1)}_{e,[m_2]}(\bm{k}_1)
\bigg[\left(q\frac{\partial}{\partial q}-1\right)f_{I,1m_1}^{(1)}(\bm{k}_2) 
C^{-,2}_{m_1 m}
\nonumber\\
&& \hspace*{0.85cm} 
- \left(q\frac{\partial}{\partial q}+4\right)
\bigg(f^{(1)}_{I,3m_1} -\sqrt{\frac{10}{3}}f_{E,3m_1}^{(1)}\bigg)(\bm{k}_2) 
C^{+,2}_{m_1 m}
+\sqrt{6}\left(q\frac{\partial}{\partial q}+1\right)
f_{B,2m_1}^{(1)}(\bm{k}_2)D^{0,2}_{m_1 m} \bigg] 
\nonumber \\
&& \hspace*{0.85cm} 
+ \,v^{(1)}_{e,[m_1]}(\bm{k}_1) v^{(1)}_{e,[m_2]}(\bm{k}_2) 
\left(q\frac{\partial f_I^{(0)}}{\partial q}
+\frac{11}{20}q^2\frac{\partial^2 f_I^{(0)}}{\partial q^2}\right) 
C^{-,2}_{m_1 m}\Bigg\}
\nonumber \\
&+& \delta_{l3} \frac{1}{10}v^{(1)}_{e,[m_2]}(\bm{k}_1)
\left(-q\frac{\partial}{\partial q}+1\right)
\left(f_{I,2m_1}^{(1)}-\sqrt{6}f_{E,2m_1}^{(1)}\right)(\bm{k}_2)
C^{-,3}_{m_1 m} \Bigg\}
\label{Imode2ndorder}\\[-0.1cm]
\nonumber
\end{eqnarray}
\begin{eqnarray}
\frac{\partial}{\partial \eta}f_{V,lm}^{(2)}(\bm{k}) &+& 
\sum_{\pm} \,(\mp i) f_{V,(l \pm 1)m_1}^{(2)}(\bm{k})\,k^{[m_2]}
C_{m_1 m}^{\pm,l}
\nonumber \\
&-& \dot{D}^{(1)}(\bm{k}_1)\,q\frac{\partial}{\partial q}
f_{V,lm}^{(1)}(\bm{k}_2) 
+ \sum_{\pm} \,(\mp i) \,\bigg\{ 
k_1^{[m_2]}\left(A^{(1)}-D^{(1)} \right)(\bm{k}_1)
f_{V,(l \pm 1)m_1}^{(1)}(\bm{k}_2) R_{m_1 m}^{\pm , l} 
\nonumber \\
&&\hspace*{0cm}
+\,\bigg(k_2^{[m_2]}\left(A^{(1)}-D^{(1)}\right)(\bm{k}_1)
-k_1^{[m_2]} A^{(1)}(\bm{k}_1)\,q\frac{\partial}{\partial q}\bigg)\,
f_{V,(l \pm 1)m_1}^{(1)}(\bm{k}_2)C_{m_1 m}^{\pm,l}\bigg\}
\nonumber \\
=|\dot{\kappa}|\,\Bigg\{ &-&f_{V,lm}^{(2)}(\bm{k}) 
+\delta_{l1}\frac{1}{2}f^{(2)}_{V,1m}(\bm{k})
\nonumber \\
&+& \bigg(A^{(1)}+\left[\frac{\delta \rho_b}{\rho_b}\right]^{(1)} + 
\left[\frac{\delta x_e}{x_e}\right]^{(1)}\,\bigg)(\bm{k}_1)
\left(-f_{V,lm}^{(1)}(\bm{k}_2) 
+\delta_{l1}\frac{1}{2}f^{(1)}_{V,1m}(\bm{k}_2)\right)
\nonumber \\[0.2cm]
&+&  \sum_{\pm}\,(\mp 1) v^{(1)}_{e,[m_2]}(\bm{k}_1) 
f_{V,(l \pm 1)m_1}^{(1)}(\bm{k}_2) C^{\pm,l}_{m_1 m}
\nonumber\\
&+&\delta_{l0}\,\frac{1}{2}v^{(1)}_{e,[m_2]}(\bm{k}_1)
\left(q\frac{\partial}{\partial q}+3\right)
f^{(1)}_{V,1m_1}(\bm{k}_2)C^{+,0}_{m_1 m}
\nonumber \\
&+&\delta_{l1}\,\frac{1}{2} v^{(1)}_{e,[m_2]}(\bm{k}_1)
\Bigg( q\frac{\partial}{\partial q}f^{(1)}_{V,0m_1}(\bm{k}_2)C^{-,1}_{m_1 m} 
- \left(q\frac{\partial}{\partial q}+3\right)f^{(1)}_{V,2m_1}(\bm{k}_2)
C^{+,1}_{m_1 m} \Bigg) 
\nonumber\\
&-& \delta_{l2} \,\frac{1}{2}v^{(1)}_{e,[m_2]}
(\bm{k}_1)q\frac{\partial}{\partial q}f^{(1)}_{V,1m_1}(\bm{k}_2)  
C^{-,2}_{m_1 m} \Bigg\} 
\\[-0.1cm]
\nonumber
\end{eqnarray}
\begin{eqnarray}
\frac{\partial}{\partial \eta}f_{E,lm}^{(2)}(\bm{k}) &+& 
\sum_{\pm} \,(\mp i) f_{E,(l \pm 1)m_1}^{(2)}(\bm{k})\,k^{[m_2]}
D_{m_1 m}^{\pm,l} - i f_{B,l m_1}^{(2)}(\bm{k}) \,k^{[m_2]} D^{0,l}_{m_1 m}  
\nonumber \\
&-& \dot{D}^{(1)}(\bm{k}_1)\,
q\frac{\partial}{\partial q}f_{E,lm}^{(1)}(\bm{k}_2) 
+ \sum_{\pm}\, (\mp i) \bigg\{ 
k_1^{[m_2]} \left(A^{(1)}-D^{(1)} \right)(\bm{k}_1)
f_{E,(l \pm 1)m_1}^{(1)}(\bm{k}_2) K_{m_1 m}^{\pm , l} 
\nonumber \\
&& +\, \bigg( k_2^{[m_2]}\left(A^{(1)}-D^{(1)}\right)(\bm{k}_1)
-k_1^{[m_2]}A^{(1)}(\bm{k}_1)q\frac{\partial}{\partial q} \bigg)
\,f_{E,(l\pm 1)m_1}^{(1)}(\bm{k}_2)D_{m_1 m}^{\pm,l}\bigg\} 
\nonumber \\[0.2cm]
&-& i k_1^{[m_2]} \left(A^{(1)}-D^{(1)} \right)(\bm{k}_1)
f_{B,lm_1}^{(1)}(\bm{k}_2) K_{m_1 m}^{0, l}  
\nonumber \\
&-&i\,\bigg( k_2^{[m_2]}\left(A^{(1)}-D^{(1)}\right)(\bm{k}_1)
-k_1^{[m_2]} A^{(1)}(\bm{k}_1)\,q\frac{\partial}{\partial q} \bigg)
\,f_{B,lm_1}^{(1)}(\bm{k}_2)D_{m_1 m}^{0,l} 
\nonumber \\
=|\dot{\kappa}|\,\Bigg\{&-& f_{E,lm}^{(2)}(\bm{k}) 
-\delta_{l2} \frac{\sqrt{6}}{10} \left(f_{I,2m}^{(2)}(\bm{k}) 
- \sqrt{6}f_{E,2m}^{(2)}(\bm{k}) \right) 
\nonumber\\
&+& 
\bigg(A^{(1)}+\left[\frac{\delta \rho_b}{\rho_b}\right]^{(1)} + 
\left[\frac{\delta x_e}{x_e}\right]^{(1)}\,\bigg)(\bm{k}_1)
\left( -f_{E,lm}^{(1)}(\bm{k}_2) -
\delta_{l2} \frac{\sqrt{6}}{10} \left(f_{I,2m}^{(1)} 
- \sqrt{6}f_{E,2m}^{(1)} \right)(\bm{k}_2)\right) 
\nonumber\\[0.2cm]
&+& \sum_{\pm} \, (\mp 1) v^{(1)}_{e,[m_2]}(\bm{k}_1) 
f_{E,(l \pm 1)m_1}^{(1)}(\bm{k}_2) D^{\pm,l}_{m_1 m} 
- v^{(1)}_{e,[m_2]}(\bm{k}_1)f_{B,lm_1}^{(1)}(\bm{k}_2) D^{0,l}_{m_1 m} 
\nonumber\\
&+& \delta_{l2} \,\frac{\sqrt{6}}{10}\,\Bigg\{v^{(1)}_{e,[m_2]}(\bm{k}_1)
\bigg[
\left(-q\frac{\partial}{\partial q}+1\right)f_{I,1m_1}^{(1)}(\bm{k}_2) 
C_{m_1 m}^{-,2}
\nonumber\\
&& \hspace*{0.85cm}
+\,\left(q\frac{\partial}{\partial q}+4\right)
\bigg(f^{(1)}_{I,3m_1}-\sqrt{\frac{10}{3}}f_{E,3m_1}^{(1)}\bigg)(\bm{k}_2) 
C^{+,2}_{m_1 m}
-\sqrt{6}\left(q\frac{\partial}{\partial q}+1\right)
f_{B,2m_1}^{(1)}(\bm{k}_2)D^{0,2}_{m_1 m}  \bigg]
\nonumber\\
\nonumber\\
&& \hspace*{0.85cm}
- \frac{1}{2}\,v^{(1)}_{e,[m_1]}(\bm{k}_1)v^{(1)}_{e,[m_2]}(\bm{k}_2) 
\,q^2\frac{\partial^2 f_I^{(0)}}{\partial q^2} C_{m_1 m}^{-,2} \bigg\} 
\nonumber \\
&+& \delta_{l3}\,\frac{\sqrt 6}{10}v^{(1)}_{e,[m_2]}(\bm{k}_1)
\left(q\frac{\partial}{\partial q}-1\right)
\left(f_{I,2m_1}^{(1)}- \sqrt{6}f_{E,2m_1}^{(1)}\right)(\bm{k}_2)
D^{-,3}_{m_1 m} \Bigg\}
\label{Emode2ndorder}\\[-0.1cm]
\nonumber
\end{eqnarray}
\begin{eqnarray}
\frac{\partial}{\partial \eta}f_{B,lm}^{(2)}(\bm{k}) &+& 
\sum_{\pm} \,(\mp i) f_{B,(l \pm 1)m_1}^{(2)}(\bm{k})\,k^{[m_2]}
D_{m_1 m}^{\pm,l} + i f_{E,l m_1}^{(2)}(\bm{k}) \,k^{[m_2]} D^{0,l}_{m_1 m}  
\nonumber \\
&-& \dot{D}^{(1)}(\bm{k}_1)\,
q\frac{\partial}{\partial q}f_{B,lm}^{(1)}(\bm{k}_2) 
+ \sum_{\pm}\, (\mp i) \bigg\{ 
k_1^{[m_2]} \left(A^{(1)}-D^{(1)} \right)(\bm{k}_1)
f_{B,(l \pm 1)m_1}^{(1)}(\bm{k}_2) K_{m_1 m}^{\pm , l} 
\nonumber \\
&& +\, \bigg( k_2^{[m_2]}\left(A^{(1)}-D^{(1)}\right)(\bm{k}_1)
-k_1^{[m_2]}A^{(1)}(\bm{k}_1)q\frac{\partial}{\partial q} \bigg)
\,f_{B,(l\pm 1)m_1}^{(1)}(\bm{k}_2)D_{m_1 m}^{\pm,l}\bigg\} 
\nonumber \\[0.2cm]
&+& i k_1^{[m_2]} \left(A^{(1)}-D^{(1)} \right)(\bm{k}_1)
f_{E,lm_1}^{(1)}(\bm{k}_2) K_{m_1 m}^{0, l}  
\nonumber \\
&+&i\,\bigg( k_2^{[m_2]}\left(A^{(1)}-D^{(1)}\right)(\bm{k}_1)
-k_1^{[m_2]} A^{(1)}(\bm{k}_1)\,q\frac{\partial}{\partial q} \bigg)
\,f_{E,lm_1}^{(1)}(\bm{k}_2)D_{m_1 m}^{0,l} 
\nonumber \\
=|\dot{\kappa}|\,\Bigg\{ &-& f_{B,lm}^{(2)}(\bm{k}) 
- \bigg(A^{(1)}+\left[\frac{\delta \rho_b}{\rho_b}\right]^{(1)} + 
\left[\frac{\delta x_e}{x_e}\right]^{(1)}\,\bigg)(\bm{k}_1)
\,f_{B,lm}^{(1)}(\bm{k}_2)
\nonumber \\[0.2cm]
&+& \sum_{\pm}\,(\mp 1) v^{(1)}_{e,[m_2]}(\bm{k}_1) 
f_{B,(l \pm 1)m_1}^{(1)}(\bm{k}_2) D^{\pm,l}_{m_1 m}
+ v^{(1)}_{e,[m_2]}(\bm{k}_1)f_{E,lm_1}^{(1)}(\bm{k}_2) D^{0,l}_{m_1 m} 
\nonumber \\
&+& \delta_{l2} \,\frac{\sqrt{6}}{10} v^{(1)}_{e,[m_2]}(\bm{k}_1)
\left(q\frac{\partial}{\partial q}+2\right)
\left(f_{I,2m_1}^{(1)}-\sqrt{6}f_{E,2m_1}^{(1)}\right)(\bm{k}_2) 
D^{0,2}_{m_1 m} \Bigg\}
\label{Bmode2ndorder}
\end{eqnarray}
These are the dynamical equations for the second-order 
photon variables. The source terms depend on products of 
first-order perturbations as well as on the second-order perturbations 
$A^{(2)}$, $D^{(2)}$, $B^{(2)}_{[m]}$, $E^{(2)}_{[m]}$ to the metric 
and to the bulk electron velocity $v_{e,[m]}^{(2)}$. To close the 
system of equations, these quantities must be determined from the 
second-order Einstein and fluid equations. 

At this point it seems appropriate to compare our results to 
those given in \cite{Pitrou:2008hy}. We already mentioned that the  
collision term in \cite{Pitrou:2008hy} takes a different form 
before expansion of the phase-space distributions around the 
equilibrium distributions, but that these structural differences 
drop out at second order, at least for the frequency-integrated 
equations. The derivation of the expanded equations in
\cite{Pitrou:2008hy} follows a different method from the one empolyed 
in the present paper by first 
considering the collision term in the electron rest frame, and then 
performing the boost to the frame, in which the electron fluid moves 
with bulk velocity $\bm{v}_e$. In contrast, we work directly in 
this frame adopting the Maxwell-Boltzmann 
distribution~(\ref{maxboltz}) for the electrons. Both methods should 
give the same results, since the Lorentz non-covariance of 
the shifted Maxwell-Boltzmann distribution is a higher-order 
effect. For a detailed comparison we note that only the 
frequency-integrated equations for the quanitites 
$\Delta^{(n)}_{X,lm}(\eta,\bm{k})$ defined in (\ref{eq:Deltadef}) 
below are given explicitly in \cite{Pitrou:2008hy} 
and that the contribution from $c_K$ in (\ref{ih6}) is 
neglected. The integrated equations can be obtained from the above 
by applying the substitution rules (\ref{eq:Deltaget}). After 
doing this we find that the structure of the equations is 
in complete agreement but we observe differences in the  
following terms: the octupole collision source term for 
$E$-mode polarization (the $\delta_{l3}$ term in our 
(\ref{Emode2ndorder})) has different numerical coefficients 
(this is corrected in the arXiv version of \cite{Pitrou:2008hy}); 
in the $B$-mode equation (our (\ref{Bmode2ndorder})) 
the coupling coefficient ${}^{\pm1}\lambda_l^m$ differs from our 
corresponding $D^{0,l}_{m\pm 1,m}$ in the 
collision term and second-order Liouville operator, and the 
terms corresponding to to the last line before the equality sign 
in (\ref{Bmode2ndorder}) are missing~\footnote{The author of 
\cite{Pitrou:2008hy} agrees with these corrections (C.~Pitrou, 
private communication).}.

\section{Discussion}
\label{sec:discussion}

\noindent 
While a numerical or even qualitative evaluation of the second-order 
Boltzmann hierarchy is beyond the scope of the present paper, we briefly 
discuss the sources of $B$-polarization contained in the equations, 
and the tight-coupling limit. Before proceeding to the discussion of
the collision term we note the different $l$ dependences in the 
weak-lensing and gravitational time-delay terms \footnote{We thank 
C.~Pitrou for drawing our attention to this point.}, which we identify
as the product terms of $A^{(1)}, D^{(1)}$, a mode momentum 
$k_1$ or $k_2$, and $f^{(1)}_{X,lm}$ 
on the left-hand side of the Boltzmann equations 
(\ref{Imode2ndorder}) -- (\ref{Bmode2ndorder}). While 
lensing of $X=I,V,E,B$ on itself is proportional to $l$ for large 
$l$, since $R^{\pm,l}, K^{\pm,l} \propto l$ (for large $l$), 
the corresponding time-delay effect is only of order $1$, since 
$C^{\pm,l}, D^{\pm,l} \propto 1$. In contrast, for conversion 
of $E$- into $B$-polarization and vice versa, weak lensing and 
time delay are effects of the same order, and both coefficients 
involved, $K^{0,l}$ and $D^{0,l}$, are only of order $1/l$ for 
large $l$.

\subsection{\boldmath $B$-mode 
polarization from scattering}

\noindent
There are two sources of $B$-mode polarization in the photon propagation 
terms on the left-hand side of the Boltzmann equations. A well-known 
mechanism is the generation of $B$ polarization when polarized radiation 
propagates through an inhomogeneous universe, usually referred to as the 
weak-lensing effect. It appears first at second-order and is contained 
in the terms involving the product of the metric perturbation $A^{(1)}$ 
or $D^{(1)}$ with the first-order $E$-mode distribution $f^{(1)}_{E,lm}$ 
in the last two lines before the equality sign in 
(\ref{Bmode2ndorder}).

$B$-mode polarization is further generated in the presence of vector or 
tensor metric perturbations. Around photon decoupling Thomson scattering 
generates the vector and tensor components of the $E$-polarization 
quadrupole which is subsequently partially converted to $B$-polarization 
through free-streaming. In the present scenario we assume that there 
are no first-order vector or tensor metric perturbations. In the absence 
of any primordial vector or tensor perturbations, they will still be 
generated at second-order, however. $B$-mode 
polarization induced by these second-order perturbations through 
free-streaming has been estimated in \cite{Mollerach:2003nq}. The 
effect turns out to be relatively small, though comparable to the 
weak-lesning effect in the small $l$-region of the BB anisotropy 
spectrum.  

The full second-order Boltzmann equations exhibit further sources 
for $B$ polarization through the collision term, which are absent 
in the first-order equation (\ref{Bmode1storder}), which contains 
only the damping term $-f_{B,lm}^{(1)}(\bm{k})$ on the right-hand 
side. The second-order collision term in (\ref{Bmode2ndorder}) contains 
products of the electron velocity and first-order intensity and 
$E$-mode perturbations. Of particular interest is the term 
\begin{equation}
\delta_{l2} \,\frac{\sqrt{6}}{10} v^{(1)}_{e,[m_2]}(\bm{k}_1)
\left(q\frac{\partial}{\partial q}+2\right) 
f_{I,2m_1}^{(1)}(\bm{k}_2) D^{0,2}_{m_1 m},
\end{equation}
which can generate a $B$-mode quadrupole directly from the 
intensity quadrupole rather than indirectly through $E$-polarization. 
A numerical analysis of the $B$-polarization generated from this 
term will be presented in \cite{BFK10}.

\subsection{Tight-coupling limit}
\label{sec:tightcoupling}

\noindent
We now examine the second-order equations in the regime where the 
electrons and photons are strongly coupled by Thomson scattering. For the 
following discussion, we are not interested in the frequency dependence 
of the photon distribution functions and integrate over $q$. We define 
the frequency-integrated multipoles
\begin{equation}\label{eq:Deltadef}
\Delta^{(n)}_{X,lm}(\eta,\bm{k}) = 
\frac{\int dq q^3 f^{(n)}_{X,lm}(\eta,\bm{k},q)}
{\int dq q^3 f^{(0)}_I(q)}.
\end{equation}
In the fluid description of photon radiation $\Delta^{(n)}_{I,00}$ 
equals the fractional perturbations of the photon number density, 
and $\Delta^{(n)}_{I,1m} = 4 v^{(n)}_{\gamma,[m]}$ is related to the 
bulk velocity of the photon fluid.

Using partial integration, derivatives on photon distributions 
can be eliminated, resulting in the following substitution 
rules in the Boltzmann equations in Section~\ref{sec:eqsummary}:
\begin{eqnarray} 
\label{eq:Deltaget}
\nonumber
f^{(0)}_I &\rightarrow& 1 \\ \nonumber
q\frac{\partial f^{(0)}_I}{\partial q} &\rightarrow& -4 \\ \nonumber
q^2 \frac{\partial^2 f^{(0)}_I}{\partial q^2} &\rightarrow& 20 \\ \nonumber
f^{(n)}_{X,lm} &\rightarrow& \Delta^{(n)}_{X,lm} \\
q\frac{\partial f^{(n)}_{X,lm}}{\partial q} 
&\rightarrow& -4 \Delta^{(n)}_{X,lm}.
\end{eqnarray}
The only term to which these rules cannot be applied is the 
$c_K$ contribution from (\ref{ih6}) to the collision term for 
$f_{I,00}^{(2)}$, which contains non-linear terms in the 
photon distribution. Inserting the Bose-Einstein distribution 
for the zeroth-order $f_I^{(0)}$ to calculate (\ref{eq:Deltadef}) 
for this term, we find 
\begin{equation}
\frac{4 q}{m_e}\,f_I^{(0)} \big(1+f_I^{(0)}\big)
+\left[\frac{4 T_e}{m_e}+\frac{q}{m_e} \,\big(1+2 f_I^{(0)}\big)\right]
q \frac{\partial f^{(0)}_I }{\partial q}
+ \frac{T_e}{m_e}\,q^2 \frac{\partial^2 f^{(0)}_I }{\partial q^2} 
\to \frac{4 (T_e-T)}{m_e}.
\end{equation}
But in the strongly coupled electron-photon plasma the electron and 
photon temperatures coincide, so this term makes no contribution to the 
frequency-integrated Boltzmann equations.

In the tight-coupling regime the collision rate $|\dot{\kappa}|$ is 
larger than any other scale of interest. The collision term drives 
the system to equilibrium, which makes the left-hand sides of the 
Boltzmann equations small. Thus the Boltzmann equations can be satisfied 
only, if the coefficients of $|\dot{\kappa}|$ in the collision term 
on the right-hand side nearly vanish. At leading order in the 
expansion in $1/|\dot{\kappa}|$ this enforces a number of relations 
among the perturbation variables. 

Looking at the first-order equations in Section~\ref{sec:eqsummary} we 
immediately find $\Delta_{V,lm}^{(1)}=\Delta_{B,lm}^{(1)}=0$ for 
all $lm$. The intensity equation (\ref{Imode1storder}) has no collision 
term for $l=0$, so the intensity monopole is unconstrained in the 
tight-coupling limit. For the dipole $l=1$, we obtain the familiar 
relation 
\begin{equation}
\Delta^{(1)}_{I,1m}(\bm{k}) = 4v_{e,[m]}^{(1)}(\bm{k}),
\label{tightcoupling1storder}
\end{equation}
which implies that the bulk velocities of the photon and electron 
plasma are equal. (The continuity equation for the electron fluid 
yields the same relation.) Continuing with the quadrupoles, we obtain from 
(\ref{Imode1storder}), (\ref{Emode1storder}) the equations 
\begin{eqnarray}
\label{eq:widewidewi}
\nonumber
\frac{9}{10}\Delta^{(1)}_{I,2m} &=& -\frac{\sqrt{6}}{10}\Delta^{(1)}_{E,2m}\\
\frac{2}{5}\Delta^{(1)}_{E,2m} &=& -\frac{\sqrt{6}}{10}\Delta^{(1)}_{I,2m},
\end{eqnarray}
which imply  $\Delta^{(1)}_{I,2m}=\Delta^{(1)}_{E,2m}=0$. Likewise, all 
higher multipoles vanish. It follows that there is no polarization in the 
tight-coupling limit, as expected, and only the intensity monopole 
and dipole are unsuppressed.

We now consider the second-order equations in the tight-coupling 
regime. It is straightforward to see that as in first order, circular 
and $B$ polarization vanish, $\Delta_{V,lm}^{(2)}=\Delta_{B,lm}^{(2)}=0$, 
as well as the multipoles higher than $l=2$ for $I$ and $E$. The 
collision term for the intensity monopole is no longer zero, 
but vanishes at leading order in the tight-coupling expansion 
after inserting the relation (\ref{tightcoupling1storder}), so the 
monopole is again unconstrained. Setting $l=1$ in (\ref{Imode2ndorder}) 
we find the tight-coupling relation
\begin{equation}
\Delta^{(2)}_{I,1m}(\bm{k}) = 4 \left(v_{e,[m]}^{(2)}(\bm{k}) + 
v_{e,[m]}^{(1)}(\bm{k}_1) \Delta^{(1)}_{I,00}(\bm{k}_2)\right),
\label{tightcoupling2ndorder}
\end{equation}
which is similar to (\ref{tightcoupling1storder}) but contains a term 
quadratic in the first-order perturbations. Finally, we examine the 
quadrupoles. For $l=2$, we can write (\ref{Imode2ndorder}), 
(\ref{Emode2ndorder}) in the form 
\begin{eqnarray}
\dot{\Delta}^{(2)}_{I,2m} + \ldots 
&=& -|\dot{\kappa}|\left[ \Delta^{(2)}_{I,2m} + P_I^{(m)} \right], 
\nonumber\\[0.2cm]
\dot{\Delta}^{(2)}_{E,2m} + \ldots 
&=& -|\dot{\kappa}|\left[ \Delta^{(2)}_{E,2m} +\sqrt{6} P_E^{(m)} \right] 
\label{psources}
\end{eqnarray}
with 
\begin{eqnarray}
P_I^{(m)}(\bm{k}) =  -\frac{1}{10}
\Big[ \Delta^{(2)}_{I,2m}(\bm{k}) 
-\sqrt{6}\Delta^{(2)}_{E,2m}(\bm{k})\Big] 
- 9 v_{e,[m_1]}(\bm{k}_1)v_{e,[m_2]}(\bm{k}_2) \,C_{m_1 m}^{-,2}
\nonumber\\[0.2cm]
P_E^{(m)}(\bm{k}) = \frac{1}{10}
\Big[ \Delta^{(2)}_{I,2m}(\bm{k}) 
-\sqrt{6}\Delta^{(2)}_{E,2m}(\bm{k}) 
-10v_{e,[m_1]}(\bm{k}_1)v_{e,[m_2]}(\bm{k}_2) \,C_{m_1 m}^{-,2}\Big]
\label{esource}
\end{eqnarray}
Setting the right-hand sides of (\ref{psources}) to zero yields the 
tight-coupling relations
\begin{eqnarray}
\Delta^{(2)}_{I,2m}(\bm{k})& =& 
10v_{e,[m_1]}(\bm{k}_1)v_{e,[m_2]}(\bm{k}_2) \,C_{m_1 m}^{-,2} ,
\nonumber\\[0.2cm]
\Delta^{(2)}_{E,2m}(\bm{k})& =& 0.
\label{tightsolutions}
\end{eqnarray}
We therefore find that there is no polarization in tight-coupling at second 
order. However, contrary to the first order, there exists a non-vanishing 
intensity quadrupole quadratic in the electron bulk 
velocity, as expected. In cartesian components and before Fourier 
transformation, Eq.~(\ref{tightsolutions}) corresponds to 
\begin{equation}
\Delta^{ij,(2)}_{I}(\bm{x}) = \frac{4}{3}\left(v_e^i(\bm{x})
v_e^j(\bm{x}) -\frac{1}{3}\delta^{ij}\, \bm{v}_e(\bm{x})^2\right)
\end{equation}
in agreement with \cite{Bartolo:2006fj}, where this result 
has been obtained from the unpolarized Boltzmann hierarchy. 
(A factor of two difference arises due to the different convention 
for expanding quantities $X$ to second order.) 
It follows from the above that the size of the quadrupole is not 
modified when the full polarized set of equations is employed, since 
the $E$-polarization quadrupole vanishes in tight coupling.

Our results are at variance, however, with \cite{Bartolo:2007dh}, 
where it has been found that the tight-coupling intensity quadrupole 
provides a large source for $B$-mode polarization. The argument is 
based on an incomplete expression for the $E$-polarization 
source term (\ref{esource}). Since the authors of \cite{Bartolo:2007dh} 
did not have the Boltzmann equations for $E$- and $B$-polarization 
at second order available, the source term without the product 
of first-order perturbations was used, 
\begin{equation}
P_E^{(m)} \to  \frac{1}{10}
\Big[ \Delta^{(2)}_{I,2m}
-\sqrt{6}\Delta^{(2)}_{E,2m}\Big] 
\to   \frac{\Delta^{(2)}_{I,2m}}{4}.
\end{equation}
The source term was  further simplified using the second of the relations 
(\ref{eq:widewidewi}) for the second-order modes as done after the 
second arrow above, which implies the assumption that there exists 
$E$-polarization in tight-coupling in contradiction with 
(\ref{tightsolutions}). We conclude  
that in this case it is clearly important that 
the full second-order polarized equations are used. Then it follows from 
(\ref{esource}) that $P_E^{(m)}=0$, and thus there is no source 
term in tight coupling that would yield $E$- and therefore $B$-polarization 
from the line-of-sight solutions of (\ref{psources}). The large 
effect reported in \cite{Bartolo:2007dh} is therefore absent. 
Polarization is only 
generated, also at second-order, once the scattering rate drops 
sufficiently so that corrections to tight coupling become relevant.

\section{Conclusion}
\label{sec:conclusion}

\noindent
In this paper we derived the complete Boltzmann hierarchy for the 
polarized photon phase-space distributions at second order 
in conformal Newtonian gauge and in the local observer rest frame 
under the assumption that vector and tensor perturbations  
are formally of second-order. This assumption is well-motivated 
by the fact that our primary aim is to study the $B$-mode 
polarization and non-gaussianity induced at second-order, when the 
primordial sources are small. A first analysis shows that the $B$-mode 
collision term contains new sources that involve the intensity 
of the perturbation rather its $E$-polarization. 
In tight-coupling we obtain the  
intensity quadrupole found earlier from the unpolarized Boltzmann 
hierarchy but no $E$-mode polarization. The equations presented 
here set the stage for their numerical evaluation, which we 
plan to present in a subsequent paper. 

\vskip0.4cm
\noindent
{\em Acknowledgements.} We thank P.~Fischer for collaboration 
during an early stage of this work, C.~Pitrou for correspondence 
on Ref.~\cite{Pitrou:2008hy} and helpful suggestions, and
K.~Klingm\"uller and Y.Y.Y. Wong for comments on the manuscript. 
M.B.~wishes to thank the 
theory groups at the University of Z\"urich and CERN for their 
hospitality during 2008 and 2009, when part of this work was 
performed. This work is supported in part by the Gottfried Wilhelm 
Leibniz programme of the Deutsche Forschungsgemeinschaft.

\appendix
\section{Summary of definitions}
\label{app:appendix}
 
\subsection{Tetrad components}
\label{app:tetrad}

\noindent 
The tetrad components before specifiying conformal Newtonian 
gauge are given to second order by
\begin{eqnarray}  
{[e_0]}^0 &=& \frac{1}{a}\,\Big(1-A + 
\frac{3}{2} {A^{(1)}}^2 - B_i^{(1)} U_i^{(1)} 
+\frac{1}{2}U_i^{(1)}U_i^{(1)}\,\Big)  
\nonumber\\[0.2cm]
{[e_0]}^i &=&  \frac{U_i}{a}
\nonumber\\[0.2cm]
{[e_k]}^0 &=&  \frac{1}{a}\left(U_k-B_k+(D^{(1)}-A^{(1)}) \,U^{(1)}_k 
+(D^{(1)}+2 A^{(1)}) \,B^{(1)}_k +
E_{kj}^{(1)} \,(U^{(1)}_j+B^{(1)}_j)
\right)
\nonumber\\[0.2cm]
{[e_k]}^i &=& \frac{1}{a}\,\Big(\delta_{ik} 
\,\Big(1-D+\frac{3}{2}{D^{(1)}}^2\,\Big) 
- E_{ik} 
+ \frac{1}{2} B_i^{(1)} B_k^{(1)} - \frac{1}{2} U_i^{(1)} U_k^{(1)} 
-  3 D^{(1)} E_{ik}^{(1)} 
\nonumber \\
&& 
- \,\frac{3}{2} E_{ij}^{(1)} E_{jk}^{(1)} 
\Big) 
\label{deftetradfull}
\end{eqnarray}
Quantities without superscript are expanded according to 
$X=X^{(1)}+X^{(2)}+\ldots$. The one simplification that has been made 
is that we set to zero the angles $\theta_k$, which defines 
the orientation of the local inertial coordinate axes relative 
to those of $x^i$. The expressions in conformal Newtonian gauge 
adopted in this paper are given in~(\ref{deftetrad}).

\subsection{Spin-weighted spherical harmonics}
\label{app:ys}

\noindent
The spin-weighted spherical harmonics are defined for 
$l\geq |s|$ and $|m|\leq l$ by 
%\begin{widetext}
\begin{eqnarray}
Y_{lm}^s(\theta,\varphi) &=& 
\left(\frac{2 l+1}{4\pi} \frac{(l+m)! (l-m)!}{(l+s)!
    (l-s)!}\right)^{\!1/2} \,
\sin^{2 l}\frac{\theta}{2} 
\nonumber\\ 
&& \times \,\sum_{r}
\left(\begin{array}{c} l-s \\ r\end{array}\right)
\left(\begin{array}{c} l+s \\ r+s-m\end{array}\right)
(-1)^{l-r-s+m} \,e^{i m\varphi} \cot^{2 r+s-m}\frac{\theta}{2}
\end{eqnarray}
such that for $s=0$ the standard spherical harmonics are recovered.  
$Y_{lm}^s$ carries spin $s$, since under a rotation of the 
coordinate system with angle $\Delta \Psi$ is transforms as
\begin{equation}
{Y'}_{lm}^s = e^{is\Delta \Psi}\,Y_{lm}^s.
\end{equation} 
For any given $s$ the  spin-weighted spherical harmonics define a 
complete set of functions on the sphere obeying the orthogonality 
relations
\begin{eqnarray}
\int d \Omega \,Y_{lm}^{s*} Y_{l'm'}^s = \delta_{l l'} \delta_{m m'}.
\label{eq:Xorthogonal}
\end{eqnarray}
Under complex conjugation  $Y_{lm}^{s*} = (-1)^{m+s} Y_{l-m}^{-s}$.
A product of two spin-weighted spherical harmonics can be combined to 
a single one using 
\begin{eqnarray}
&&Y_{l_1 m_1}^{s_1} Y_{l_2 m_2}^{s_2} = 
\sum \limits_{l,m,s}\frac{\sqrt{(2l_1+1)(2l_2+1)}}{\sqrt{4\pi(2l+1)}} 
\,\left(\begin{array}{ccc} l_1 & l_2 & l \\ m_1 & m_2 &
    m\end{array}\right) 
\left(\begin{array}{ccc} l_1 & l_2 & l \\ -s_1 & -s_2 &
    -s\end{array}\right) 
Y_{lm}^s\qquad
 \label{eq:Xcgrelation} 
\end{eqnarray}
The summation ranges are restricted by the triangular equation for 
the Clebsch-Gordan coefficients 
\begin{equation}\label{eq:XDreieckCGrelation}
\left(\begin{array}{ccc}l_1 & l_2 & l \\m_1 & m_2 &
    m\end{array}\right) \not= 0 
\ \ \text{if}  \ \ |l_2-l_1| \leq l \leq l_1+l_2, 
\end{equation}
and
\begin{equation}
\label{eq:Xmkonservation}
\left(\begin{array}{ccc}l_1 & l_2 & l \\m_1 & m_2 &
    m\end{array}\right)= 0 
\ \ \text{if}  \ \ m \neq m_1+m_2.
\end{equation}
This implies in particular $s=s_1+s_2$ in (\ref{eq:Xcgrelation}). 
Furthermore, the Clebsch-Gordan coefficients satisfy the  
following relation:
\begin{equation}
\label{eq:Xmflip}
\left(\begin{array}{ccc}l_1 & l_2 & l \\m_1 & m_2 &
    m\end{array}\right)=
(-1)^{l_1+l_2-l}\left(\begin{array}{ccc}l_1 & l_2 & l \\-m_1 & -m_2 &
    -m
\end{array}\right).
\end{equation}

The spin-raising and -lowering operators are 
defined by 
\begin{eqnarray}
\eth_s &=& -\frac{\partial}{\partial \theta} 
-\frac{i}{\sin\theta}\frac{\partial}{\partial \varphi} + s \cot\theta,
\nonumber\\
\bar{\eth}_s &=& -\frac{\partial}{\partial \theta} +
\frac{i}{\sin\theta}\frac{\partial}{\partial \varphi} - s \cot\theta.
\label{thop}
\end{eqnarray}
We then have
\begin{equation}
\eth_s Y_{lm}^s = \phantom{-}l_s^+ Y_{lm}^{s+1}, 
\quad
\bar{\eth}_s Y_{lm}^s = -l_s^- Y_{lm}^{s-1},
\label{derYs}
\end{equation}
where
\begin{equation}
l_s^\pm = \sqrt{(l\mp s)(l\pm s+1)}.
\end{equation}

\subsection{Unit vector in the spherical basis}
\label{xichi}

\noindent The coefficients $\xi_m^i$ and $\chi^{ij}_{2 m}$ defined in 
(\ref{sphericaln}) which 
express $n^i$ and $n^i n^j$ in terms of spherical harmonics 
are given explicitly by:
\begin{equation}
\xi_0 = \left(\begin{array}{c}
0\\0\\1
\end{array}\right)
\quad\qquad
\xi_{\pm1} = \frac{1}{\sqrt{2}}\left( \begin{array}{c}
\mp 1\\i\\0
\end{array}\right),
\end{equation}
\begin{equation}
\chi_{20} = \frac{1}{3}\left( \begin{array}{ccc}
-1 & 0 & 0 \\0 & -1 & 0\\0 & 0 & 2 
        \end{array}
\right)\qquad
\chi_{2,\pm 1} = \frac{1}{\sqrt{6}}\left( \begin{array}{ccc}
0&0&\mp 1\\0&0&i\\ \mp 1&i&0
\end{array}\right)
\qquad
\chi_{2,\pm2} = \frac{1}{\sqrt{6}}\left( \begin{array}{ccc}
1&\mp i&0\\ \mp i& -1& 0\\  0&0&0
\end{array}\right).
\end{equation}

\subsection{\boldmath $Q$ matrices and $\omega$ coefficients}

\noindent
The non-vanishing $Q$-matrices introduced in (\ref{QmatricesIVEB}) 
read in the $IVEB$ basis:
\begin{equation} 
\label{eq:XQsexplizit}
\begin{array}{ll}
Q_{I,00}^{ij}=\displaystyle \frac{1}{3}
\left(\begin{array}{ccc}1 & 0 & 0 \\0 & 1 & 0 \\0 & 0 & 1\end{array}\right) 
\phantom{some more space}
&
\\[0.8cm]
Q_{I,20}^{ij}=\displaystyle \frac{1}{6\sqrt{5}}
\left(\begin{array}{ccr}1 & 0 & 0 \\0 & 1 & 0 \\0 & 0 & -2\end{array}\right) 
&
\\[0.8cm]
Q_{I,21}^{ij}=\displaystyle\frac{1}{2\sqrt{30}}
\left(\begin{array}{ccc}0 & 0 & 1 \\0 & 0 & i \\1 & i & 0\end{array}\right) 
&
Q_{I,2-1}^{ij}=\displaystyle\frac{1}{2\sqrt{30}}
\left(\begin{array}{rcr}0 & 0 & -1 \\0 & 0 & i \\-1 & i & 0\end{array}\right) 
\\[0.8cm]
Q_{I,22}^{ij}=\displaystyle\frac{1}{2\sqrt{30}}
\left(\begin{array}{rrc}-1 & -i & 0 \\-i & 1 & 0 \\0 & 0 & 0\end{array}\right) 
&
Q_{I,2-2}^{ij}=\displaystyle\frac{1}{2\sqrt{30}}
\left(\begin{array}{rcc}-1 & i & 0 \\i & 1 & 0 \\0 & 0 & 0\end{array}\right)
\\[0.8cm]
Q_{V,10}^{ij}=\displaystyle\frac{1}{2\sqrt{3}}
\left(\begin{array}{crc}0 & -i & 0 \\i & 0 & 0 \\0 & 0 & 0\end{array}\right) 
&
\\[0.8cm]
Q_{V,11}^{ij}=\displaystyle\frac{1}{2\sqrt{6}}
\left(\begin{array}{crc}0 & 0 & 1 \\0 & 0 & i \\ -1 & -i & 0\end{array}\right) 
&
Q_{V,1-1}^{ij}=\displaystyle\frac{1}{2\sqrt{6}}
\left(\begin{array}{rcr}0 & 0 & 1 \\0 & 0 & -i \\-1 & i & 0\end{array}\right)
\\[0.8cm]
Q_{E,20}^{ij}=\displaystyle\frac{1}{\sqrt{30}}
\left(\begin{array}{rrc}-1 & 0 & 0 \\0 & -1 & 0 \\0 & 0 & 2\end{array}\right) 
&
\\[0.8cm]
Q_{E,21}^{ij}=\displaystyle\frac{1}{2\sqrt{5}}
\left(\begin{array}{rrr}0 & 0 & -1 \\0 & 0 & -i \\-1 & -i & 0
\end{array}\right)
&
Q_{E,2-1}^{ij}=\displaystyle\frac{1}{2\sqrt{5}}
\left(\begin{array}{crr}0 & 0 & 1 \\0 & 0 & -i \\1 & -i & 0\end{array}\right)
\\[0.8cm]
Q_{E,22}^{ij}=\displaystyle\frac{1}{2\sqrt{5}}
\left(\begin{array}{crc}1 & i & 0 \\i & -1 & 0 \\0 & 0 & 0\end{array}\right) 
&
Q_{E,2-2}^{ij}=\displaystyle\frac{1}{2\sqrt{5}}
\left(\begin{array}{rrc}1 & -i & 0 \\-i & -1 & 0 \\0 & 0 & 0\end{array}\right)
\end{array} 
\end{equation}

The non-zero traces (\ref{wsymbols}) are:
\begin{eqnarray}
\label{wsymbolvalues}
&& \omega^{(0)}_{II}=1, 
\nonumber\\[0.2cm]
&&\omega^{(1)}_{VV}=\frac{1}{2}, 
\nonumber\\
&& \omega^{(2)}_{II}=\frac{1}{10},
\qquad \omega^{(2)}_{EE}= \frac{3}{5}, 
\qquad
\omega^{(2)}_{IE}=\omega^{(2)}_{EI}= -\sqrt{\frac{3}{50}}. 
\end{eqnarray}
\label{ch:bessel}

%\bibliography{cosmo}

\end{document}